\documentclass[12pt]{article}

\usepackage{amsmath,amssymb,epsfig,subfigure,color,soul,fleqn,rotating}
\usepackage{ulem,slashed}
\usepackage{cite}
\usepackage[colorlinks,citecolor=blue,urlcolor=blue,linkcolor=blue]{hyperref}
\usepackage{xcolor}

\newcommand{\be}{\begin{eqnarray}}
\newcommand{\ee}{\end{eqnarray}}
\newcommand{\nee}{\nonumber\end{eqnarray}}

\newcommand{\msbar}{{\overline{\rm MS}}}
\newcommand{\drbar}{{\overline{\rm DR}}}

\newcommand{\mch}[1] {m_{\ti \x^+_{#1}}}
\newcommand{\mnt}[1] {m_{\ti \x^0_{#1}}}

\newcommand{\msg}    {m_{\ti g}}
\newcommand{\msu}[1] {m_{\ti u_{#1}}}
\newcommand{\msd}[1] {m_{\ti d_{#1}}}

\newcommand{\lsim}{\;\raisebox{-0.9ex}
          {$\textstyle\stackrel{\textstyle<}{\sim}$}\;}

\def\be            {\begin{equation}}
\def\ee            {\end{equation}}
\def\bea            {\begin{eqnarray}}
\def\eea            {\end{eqnarray}}

\definecolor{mybrown}{cmyk}{0,0.9,1.5,0.3}

\def\a              {\alpha}
\def\b               {\beta}
\def\d               {\delta}
\def\g               {\gamma}

\def\x               {\chi}
\def\ti              {\tilde}
\def\sq              {\ti q}
\def\st              {\ti t}
\def\sc              {\ti c}
\def\sb              {\ti b}
\def\ch              {\ti \x^\pm}

\def\nt              {\ti \x^0}

\def\stau            {\ti \tau}

\def\su                {\ti{u}}

\def \sca                 {\ti{c}}
\def\sd                {\ti{d}}

\newcommand{\AddrGAKUGEI}{%
 \it Department of Physics, Tokyo Gakugei University, Koganei,
Tokyo 184-8501, Japan\\}

\newcommand{\AddrHEPHY}{%
 \it Institut f\"ur Hochenergiephysik der \"Osterreichischen Akademie
der Wissenschaften, A-1050 Vienna, Austria\\}

\newcommand{\AddrElena}{%
 \it VRVis Zentrum f\"ur Virtual Reality und Visualisierung Forschungs-GmbH,  A-1220 Vienna, Austria\\}

\title{\bf Correlation between the decays \boldmath $h^0 \to \gamma \gamma / g g$ 
in the MSSM with quark flavour violation }

\author{H. Eberl${}^{1}$, K.~Hidaka${}^{2}$, E. Ginina${}^{1,3}$}

\date{
\small $^1$ \AddrHEPHY
          $^2$ \AddrGAKUGEI
          $^3$ \AddrElena
}

\definecolor{darkgreen}{rgb}{0,.5,0}

\begin{document}

\begin{flushright}
HEPHY-PUB 1013/18\\
\end{flushright}

\begingroup
\let\newpage\relax
\maketitle
\endgroup

\maketitle
\thispagestyle{empty}

\begin{abstract}
We study the loop-induced decays $h^0 \to \gamma \, \gamma$ and 
$h^0 \to g \, g$ in the Minimal Supersymmetric Standard Model (MSSM) with quark flavour 
violation (QFV), identifying $h^0$ with the Higgs boson with a mass of 125 GeV, where $\gamma$ 
and $g$ are photon and gluon, respectively. We perform a MSSM parameter scan 
and a detailed analysis around a fixed reference point respecting 
theoretical constraints from vacuum stability conditions and experimental constraints, such 
as those from B meson data and electroweak precision data, as well as recent limits on supersymmetric 
(SUSY) particle masses from LHC experiments. We find that 
(i) the relative deviation of the decay width $\Gamma(h^0 \to g \, g)$ from 
the Standard Model value, $DEV(g)$, can be large and negative, $\lsim - 15\%$,
(ii) the analogous deviation of $\Gamma(h^0 \to \gamma \, \gamma)$ is strongly correlated,
$DEV(\gamma) \simeq -1/4\,DEV(g)$ for $DEV(g) \lsim - 4\%$,
(iii) the relative deviation of the width ratio $\Gamma(h^0 \to \gamma \, \gamma)/\Gamma(h^0 \to g \, g)$
from the SM value, $DEV(\gamma/g)$, can be large (up to $\sim$ 20\%),
(iv) the deviations can be large due to the up-type squark loop contributions,
(v) the SUSY QFV parameters can have a significant effect on these deviations. 
Such large deviations can be observed at a future $e^+e^-$ collider like ILC. 
Observation of the deviation patterns as shown in this 
study would favour the MSSM with flavour-violating squark mixings 
and encourage to perform further studies in this model. 
\end{abstract}

\clearpage

\section{Introduction}

The Standard Model (SM) is a very successful theory of elementary particle physics. 
It is, however, known to have several essential problems. 
Primarily it fails to provide an explanation of observed phenomena like the 
neutrino masses, the matter-antimatter asymmetry, and the dark matter origin. 
Therefore, it is necessary to search for New Physics, that will help to complete 
the theory, solve its problems and account the missing details.

Recently a Higgs boson with mass of 125 GeV has been discovered at the Large Hadron Collider (LHC) 
~\cite{Aad:2012tfa, Chatrchyan:2012xdj} that behaves like the Higgs boson of the SM. 
Whether it is indeed the SM Higgs boson or a Higgs boson of New Physics beyond the SM, this is 
presently one of the most important issues in particle physics.  A detailed study of the 
properties of the Higgs boson can provide a crucial clue in the search for the 
ultimate New Physics theory. The theory of Supersymmetry (SUSY) is the most prominent 
candidate for a New Physics theory solving the SM problems. In this paper we study the 
possibility that the discovered Higgs boson is the lightest CP-even neutral Higgs boson 
$h^0$ of the Minimal Supersymmetric Standard Model (MSSM) \cite{LHCcrosssecs, Djouadi:2005gi}. 

In the phenomenological analysis of the MSSM, quark flavour conservation (QFC) is 
usually assumed, apart from the quark flavour violation (QFV) induced by the 
Cabibbo-Kobayashi-Maskawa matrix. However, SUSY QFV terms could be present 
in the mass matrix of the squarks. Especially important can be the mixing terms between the 2nd and 
the 3rd squark generations, such as $\sca_{L,R}-\st_{L,R}$ mixing terms, where 
$\sca$ and $\st$ are the charm- and top-squark, respectively.

In ~\cite{Bartl:2014bka} we pointed out the importance of 
the SUSY QFV effects due to squark loop contributions in the decays of the MSSM 
Higgs boson $h^0$. We showed that the QFV effect due to $\sca_{L,R}-\st_{L,R}$ mixing  
can have a major impact on the decay $h^0 \to c \, \bar{c}$,  strongly enhancing
the deviation of the MSSM Higgs boson decay rate $\Gamma(h^0 \to c \, \bar{c})$ 
from the SM Higgs boson decay rate $\Gamma(H_{SM} \to c \, \bar{c})$, where c is the charm-quark.
In \cite{Eberl:h2bb} we also showed that the QFV due to $\sca_{L,R}-\st_{L,R}$ 
mixing can significantly enhance the difference between $\Gamma(h^0 \to b \, \bar{b})$ and 
$\Gamma(H_{SM} \to b \, \bar{b})$, where b is the bottom-quark.

The loop-induced decays $h^0 \to \gamma \, \gamma$ and $h^0 \to g \, g$ are very sensitive 
to New Physics since loops of New Physics particles can appear at the lowest order of 
perturbative expansion of the decay amplitudes.
The rates of these loop-induced decays 
were already calculated including gluonic QCD \cite{QCD_corr} and electroweak
\cite{EW_corr} radiative corrections in the SM and also partly in the MSSM
with QFC (except \cite{Brignole:2015kva} mentioned below).
In this paper we study the influence of the SUSY QFV due to
$\sca_{L,R}-\st_{L,R}$ mixing on $h^0 \to \gamma \, \gamma$ 
and $h^0 \to g \, g$, including the gluonic two-loop QCD corrections \cite{Spira}. 
(We also studied $\ti{s}_{L,R}-\ti{b}_{L,R}$ mixing, with $\ti{s}$ and $\ti{b}$ the strange- 
and bottom-squark, respectively,
but the effects turned out to be very small.)
For this purpose, we perform a MSSM parameter scan respecting theoretical constraints 
from vacuum stability conditions and experimental constraints, such as those from 
B meson data and electroweak precision data, as well as recent limits on SUSY particle 
masses from LHC experiments. 
In \cite{Brignole:2015kva} these loop-induced decays were studied in the MSSM with QFV
in an effective field theory approach based on dim-6 operators in a so-called 
$\kappa$-framework. 
However, that paper does not take into account the radiative
corrections and the constraints mentioned above, except those from the electroweak
precision data. Moreover, it does not include the $\sc_{R}-\st_{R}$ mixing effect. 
As we will point out later, this mixing effect can also play an important role in 
the considered loop-induced decays.

Although the $h^0$ decay widths of the $\gamma \gamma$ and $g g$ modes are studied in the SM and 
the MSSM in many articles \cite{QCD_corr} - \cite{Nojiri_Boselli}, a systematic  
numerical study of the deviations of the MSSM widths from the SM values taking into 
account the SUSY QFV effect and the constraints is still missing. 
In this article we thoroughly perform such a study with 
special emphasis on the importance of SUSY QFV. Furthermore, we elucidate 
the sensitivities of measurements at the LHC and at future lepton colliders, such as ILC, to the deviations.
 
As lepton-flavour violation effect has turned out to be very small
in our analysis, we assume lepton flavour conservation.
We also assume that the lightest neutralino is the lightest SUSY particle (LSP).

In the following section we introduce the SUSY QFV parameters originating from the squark mass matrices. 
Details about our parameters scan are given in Section~\ref{sec:full scan}. In Section~\ref{sec:correlation} 
we define the deviations of the widths $h^0 \to \gamma \, \gamma$ and $h^0 \to g \, g$ from the SM and 
analyse their behaviour in the  studied SUSY QFV scenarios. The paper rounds up with conclusions, 
contained in Section~\ref{sec:concl}, and one short Appendix, where all relevant constraints are listed.

%
\section{Squark mass matrices in the MSSM with flavour violation}
\label{sec:sq.matrix}
%
In the super-CKM basis of $\sq_{0 \gamma} =
(\sq_{1 {\rm L}}, \sq_{2 {\rm L}}, \sq_{3 {\rm L}}$,
$\sq_{1 {\rm R}}, \sq_{2 {\rm R}}, \sq_{3 {\rm R}}),~\gamma = 1,...6,$  
with $(q_1, q_2, q_3)=(u, c, t),$ $(d, s, b)$, the up-type and down-type squark mass matrices 
${\cal M}^2_{\tilde{q}},~\tilde{q}=\tilde{u},\tilde{d}$, at the SUSY scale have the following 
most general $3\times3$ block form~\cite{Allanach:2008qq}:
\begin{equation}
    {\cal M}^2_{\tilde{q}} = \left( \begin{array}{cc}
        {\cal M}^2_{\tilde{q},LL} & {\cal M}^2_{\tilde{q},LR} \\[2mm]
        {\cal M}^2_{\tilde{q},RL} & {\cal M}^2_{\tilde{q},RR} \end{array} \right), \quad \tilde{q}=\tilde{u},\tilde{d}\,.
 \label{EqMassMatrix1}
\end{equation}
Non-zero off-diagonal terms of the $3\times3$ blocks ${\cal M}^2_{\tilde{q},LL},~{\cal M}^2_{\tilde{q},RR},~{\cal M}^2_{\tilde{q},LR}$ 
and ${\cal M}^2_{\tilde{q},RL}$ in Eq.~(\ref{EqMassMatrix1}) explicitly break the quark-flavour in the squark sector of the MSSM.
The left-left and right-right blocks in Eq.~(\ref{EqMassMatrix1}) are given by
\begin{eqnarray}
    & &{\cal M}^2_{\tilde{u}(\tilde{d}),LL} = M_{Q_{u(d)}}^2 + D_{\tilde{u}(\tilde{d}),LL}{\bf 1} + \hat{m}^2_{u(d)}, \nonumber \\
    & &{\cal M}^2_{\tilde{u}(\tilde{d}),RR} = M_{U(D)}^2 + D_{\tilde{u}(\tilde{d}),RR}{\bf 1} + \hat{m}^2_{u(d)},
     \label{EqM2LLRR}
\end{eqnarray}
where $M_{Q_{u}}^2=V_{\rm CKM} M_Q^2 V_{\rm CKM}^{\dag}$, $M_{Q_{d}}^2 \equiv M_Q^2$, 
$M_{Q,U,D}$ are the hermitian soft SUSY-breaking mass matrices of the squarks, 
$D_{\tilde{u}(\tilde{d}),LL}$, $D_{\tilde{u}(\tilde{d}),RR}$ are the $D$-terms, and  
$\hat{m}_{u(d)}$ are the diagonal mass matrices of the up(down)-type quarks.
$M_{Q_{u}}^2$ is related with $M_{Q_{d}}^2$
by the CKM matrix $V_{\rm CKM}$ due to the $SU(2)_{\rm L}$ symmetry.
The left-right and right-left blocks of Eq.~(\ref{EqMassMatrix1}) are given by
\begin{eqnarray}
 {\cal M}^2_{\tilde{u}(\tilde{d}),RL} = {\cal M}^{2\dag}_{\tilde{u}(\tilde{d}),LR} &=&
\frac{v_2(v_1)}{\sqrt{2}} T_{U(D)} - \mu^* \hat{m}_{u(d)}\cot\beta(\tan\beta),
\label{M2sqdef}
\end{eqnarray}
where $T_{U,D}$ are the soft SUSY-breaking trilinear 
coupling matrices of the up-type and down-type squarks entering the Lagrangian 
${\cal L}_{int} \supset -(T_{U\alpha \beta} \ti{u}^\dagger _{R\a}\ti{u}_{L\b}H^0_2 $ 
$+ T_{D\alpha \beta} \ti{d}^\dagger _{R\a}\ti{d}_{L\b}H^0_1)$,
$\mu$ is the higgsino mass parameter, and 
$\tan\beta = v_2/v_1$ with $v_{1,2}=\sqrt{2} \left\langle H^0_{1,2} \right\rangle$.
The squark mass matrices are diagonalized by the $6\times6$ unitary matrices $U^{\tilde{q}}$,
$\tilde{q}=\tilde{u},\tilde{d}$, such that
\begin{eqnarray}
&&U^{\tilde{q}} {\cal M}^2_{\tilde{q}} (U^{\tilde{q} })^{\dag} = {\rm diag}(m_{\tilde{q}_1}^2,\dots,m_{\tilde{q}_6}^2)\,,
\label{Umatr}
\end{eqnarray}
with $m_{\tilde{q}_1} < \dots < m_{\tilde{q}_6}$.
The physical mass eigenstates
$\sq_i, i=1,...,6$ are given by $\sq_i =  U^{\sq}_{i \alpha} \sq_{0\alpha} $.

In this paper we focus on the $\ti{c}_L - \ti{t}_L$, $\ti{c}_R - \ti{t}_R$, $\ti{c}_R - \ti{t}_L$, and 
$\ti{c}_L - \ti{t}_R$ mixing which is described by the QFV parameters $M^2_{Q23}$, $M^2_{U23}$, $T_{U23}$ 
and $T_{U32}$, respectively. 
 We will also often refer to the QFC parameter $T_{U33}$ which induces the $\ti{t}_L - \ti{t}_R$ mixing 
and plays an important role in this study.\\
The slepton parameters are defined analogously to the squark ones. 
All the parameters in this study are assumed to be real, except the 
CKM matrix $V_{CKM}$.

\section{Parameter scan}
\label{sec:full scan}

We perform a MSSM parameter scan taking into account theoretical constraints 
from vacuum stability conditions and experimental constraints from K- 
and B-meson data, the $h^0$ mass and coupling data and electroweak precision 
data, as well as limits on SUSY particle masses from recent LHC experiments 
(see Appendix A). 
As for the squark generation mixings, we only consider the 
mixing between the second and third generation of squarks. The mixing between 
the first and the second generation squarks is very strongly constrained by the 
K and D meson data ~\cite{Gabbiani:1996hi, PDG2016}. 
The experimental constraints on the mixing of first and third generation squarks 
are not so strong ~\cite{Dedes}, but we don't consider this mixing since its 
effect is essentially similar to that of the mixing of second and third 
generation squarks. The parameter points are generated by using random numbers 
in the ranges shown in Table~\ref{table1}, some parameters are fixed (given in 
the last box). All parameters are defined at scale Q = 1 TeV, except $m_A(pole)$ 
which is the pole mass of the CP odd Higgs boson $A^0$.
The parameters that are not shown explicitly are taken to be zero. 
The entire scan lies in the decoupling Higgs limit, i.e. in the scenarios 
with large $\tan\beta \geq 10$ and large $m_A \geq 800$ GeV (see Table~\ref{table1}), respecting 
the fact that the discovered Higgs boson is SM-like. It is well known that the 
lightest MSSM Higgs boson $h^0$ is SM-like (including its couplings) in this limit.
Note that we don't assume the GUT relation for the gaugino masses $M_1$, $M_2$, $M_3$.
\begin{table}[h!]
\footnotesize{
\caption{
Scanned ranges and fixed values of the MSSM parameters (in units of GeV or GeV$^2$, 
except for $\tan\beta$). $M_{1,2,3}$ are the U(1), SU(2), SU(3) gaugino mass parameters.}
\begin{center}
\begin{tabular}{|c|c|c|c|c|c|}
    \hline
\vspace*{-0.3cm}
& & & & &\\
\vspace*{-0.3cm}
     $\tan\beta$ & $M_1$ &  $M_2$ & $M_3$ & $\mu$ &  $m_A(pole)$\\ 
& & & & &\\
    \hline
\vspace*{-0.3cm}
& & & & &\\
\vspace*{-0.3cm}
     10 $\div$ 30 & $100 \div 2500$ & $100 \div 2500$  & $2500 \div 5000$ & $100 \div 2500$ & $800 \div 3000$\\
& & & & &\\
    \hline
    \hline
\vspace*{-0.3cm}
& & & & &\\
\vspace*{-0.3cm}
      $ M^2_{Q 22}$ & $ M^2_{Q 33}$ &  $|M^2_{Q 23}| $ & $ M^2_{U 22} $ & $ M^2_{U 33} $ &  $|M^2_{U 23}| $\\ 
& & & & &\\
     \hline
\vspace*{-0.3cm}
& & & & &\\
\vspace*{-0.3cm}
      $2500^2 \div 4000^2$ & $2500^2 \div 4000^2$ & $< 1000^2$  & $1000^2 \div 4000^2$ & $600^2 \div 3000^2$& $ < 1200^2$\\
& & & & &\\
    \hline
    \hline
\vspace*{-0.3cm}    
& & & & &\\
\vspace*{-0.3cm}      
      $ M^2_{D 22} $ & $ M^2_{D 33}$ &  $ |M^2_{D 23}|$ & $|T_{U 23}|  $ & $|T_{U 32}|  $ &  $|T_{U 33}|$\\ 
& & & & &\\
    \hline
\vspace*{-0.3cm}      
& & & & &\\
\vspace*{-0.3cm}  
       $ 2500^2 \div 4000^2$ & $1000^2 \div 3000^2 $ & $ < 1000^2$  & $< 4000 $ & $ < 4000$& $< 4000 $\\
& & & & &\\
 \hline 
\multicolumn{6}{c}{}\\[-3.6mm]  
\cline{1-4}
\vspace*{-0.3cm}      
     & & & \\
\vspace*{-0.3cm}      
     $ |T_{D 23}| $ & $|T_{D 32}|  $ &  $|T_{D 33}|$ &$|T_{E 33}| $\\ 
     & & & \\
    \cline{1-4}
\vspace*{-0.3cm}      
     & & & \\
\vspace*{-0.3cm}      
     $< 1000 $ & $< 1000 $& $ < 1000$& $ < 500$\\
     & & & \\
    \cline{1-4}
\end{tabular}\\[3mm]
\begin{tabular}{|c|c|c|c|c|c|c|c|c|}
    \hline
\vspace*{-0.3cm}      
    & & & & & & & &\\
\vspace*{-0.3cm}      
    $M^2_{Q 11}$ & $M^2_{U 11} $ &  $M^2_{D 11} $ & $M^2_{L 11}$ & $M^2_{L 22} $ &  $M^2_{L 33}$ & $M^2_{E 11}$&$M^2_{E 22}$ & $M^2_{E 33} $\\ 
    & & & & & & & &\\
    \hline
\vspace*{-0.3cm}      
    & & & & & & & &\\
\vspace*{-0.3cm}      
    $4500^2$ & $4500^2$ & $4500^2$  & $1500^2$ & $1500^2$ & $1500^2$& $1500^2$& $1500^2$&$1500^2$\\
    & & & & & & & &\\
    \hline
\end{tabular}
\end{center}
\label{table1}
}
\end{table}

The decay widths $\Gamma(h^0 \to \gamma \gamma)_{MSSM}$ and $\Gamma(h^0 \to g g)_{MSSM}$
are calculated with our own code based on the public code {\tt SPheno}~\cite{SPheno1,SPheno2}. 
For the calculation of the MSSM spectrum we use the version {\tt SPheno-v3.3.8}. The computation 
includes lowest order 1-loop contributions and gluonic 2-loop QCD corrections (i.e. NLO QCD corrections) 
to quark loops \cite{Spira}~\footnote{
The gluonic 2-loop QCD corrections to the squark loops are negligibly small since the squark-loop contributions 
to the widths are rather small due to large squark masses from the LHC limit (see Appendix A). 
As the corrections to small contributions are very small, we can neglect such corrections. 
We can also neglect SUSY-QCD corrections to the quark/squark-loops since gluino/squarks are required to be 
so heavy by the LHC limits (see Appendix A) that gluino/squark-loop corrections (i.e. SUSY-QCD corrections) 
to the widths are very small. 
Moreover, the NNLO QCD corrections \cite{QCD_corr} and the NLO electroweak (EW) corrections \cite{EW_corr} 
to the widths are found to be much smaller than the NLO QCD corrections. 
Therefore, we take into account only the gluonic 2-loop QCD corrections (i.e. NLO QCD corrections) to quark-loop 
contributions to $\Gamma(h^0 \to \gamma \gamma / g g)_{MSSM}$.
}. 
The lowest order 1-loop contributions to $\Gamma(h^0 \to \gamma \gamma)_{MSSM}$ 
stem from the loops with SM particles, quarks (t, b, ...), charged leptons 
($\tau^-$, ...) and $W^\pm$ boson and SUSY particles, squarks ($\su$, $\sd$), 
charged sleptons ($\stau^-$, ...), charginos $\ch$ and charged Higgs bosons $H^\pm$. 
The lowest order 1-loop contributions to $\Gamma(h^0 \to g g)_{MSSM}$ 
stem from the loops with quarks (t, b, ...) and squarks ($\su$, $\sd$).
In order to stay consistent we also use our own code for the SM decay widths 
$\Gamma(h^0 \to \gamma \gamma)_{SM} \equiv \Gamma(H_{SM} \to \gamma \gamma)$, and 
$\Gamma(h^0 \to g g)_{SM} \equiv \Gamma(H_{SM} \to g g)$ including the gluonic 2-loop 
QCD corrections \cite{Spira}. We have cross-checked them numerically 
with the decoupling limit of the MSSM results.
The Higgs mass in the kinematic factors of the widths is
fixed by the measured mass at LHC, $m_{h^0} = 125.09$~GeV to avoid an artificially large dependence
stemming from the kinematic factor in $\Gamma(h^0 \to \gamma \gamma/g g)_{MSSM}$ , which is proportional 
to $m^3_{h^0}$. All MSSM input parameters are taken as $\drbar$ parameters at the scale $Q = 1$~TeV and 
then transformed by RGEs to those at the scale of $Q = m_{h^0} = 125.09$~GeV. The masses and 
rotation matrices of the sfermions are renormalized at one-loop level within SPheno based on 
the technique given in \cite{Pierce}.

From 2850000 input points generated in the scan about 285500 survived all constraints. 
These are about 10\%. We show these survival points in all scatter plots in this article.
 
\section{Deviation of the MSSM widths from the SM}
\label{sec:correlation}

We define the relative deviation of the MSSM width from the SM width 
as\footnote{For reference, the SM predictions (at 68\% CL) of \cite{Almeida:2013jfa} are
$\Gamma(\gamma)_{SM} = (1.08 ^{+0.03}_{-0.02}) \cdot 10^{-5}$ GeV and
$\Gamma(g)_{SM} = (3.61 \pm 0.06) \cdot 10^{-4}$ GeV, and those of \cite{CERN_YR4} are
$\Gamma(\gamma)_{SM} = (9.31 \pm 0.09) \cdot 10^{-6}$ GeV and                                     
$\Gamma(g)_{SM} = (3.35 \pm 0.21) \cdot 10^{-4}$ GeV.}    
\be
    DEV(X) = \Gamma(h^0 \to X X)_{MSSM}/\Gamma(h^0 \to X X)_{SM} - 1\, ,   \,   \mbox{with } X = \gamma, g\, ,
\ee
where we identify $h^0$ with the Higgs boson with a mass of 125.09 GeV.

\noindent
The relative deviation of the width ratio from the SM prediction is defined as
\be
    DEV(\gamma/g) = [\Gamma(\gamma)/\Gamma(g)]_{MSSM}/[\Gamma(\gamma)/\Gamma(g)]_{SM}  - 1
    \label{eq_DEV(ga/g)}
\ee
with 
\be
\Gamma(X) = \Gamma(h^0 \to X X), \mbox{where} \, X = \gamma, g.
\ee

\noindent
Note that $DEV(\gamma/g)$ in Eq.~(\ref{eq_DEV(ga/g)}) can be written also directly in terms of $DEV(\gamma)$ and $DEV(g)$,
\begin{equation}
DEV(\gamma/g) = {DEV(\gamma) + 1 \over DEV(g) + 1} - 1\, .
 \label{eq_DEV(X/Y)_1}
\end{equation}

Before we show the results of the full parameter scan, we briefly
comment on an expected qualitative behaviour of $DEV(g)$.
One can approximate $DEV(g)$ in an effective field theory approach based on dim-6 operators parametrized in 
a so-called $\kappa$-framework~\cite{Brignole:2015kva}, assuming that the SM contribution stems only from 
the top-loop and neglecting the Higgs mass in the amplitude.
Based on the result for $\delta \kappa_g$ given in~\cite{Brignole:2015kva}, we can write the approximation for 
$DEV(g) \sim 2 \delta \kappa_g = DEV(g)^{approx}$  in our convention (see Section \ref{sec:sq.matrix}),
\begin{equation}
DEV(g)^{approx} =  {v^2 \over 4} \Bigg[ {1 \over m^2_{\st_L} } \left( y_t^2  - {|T_{U23}|^2 \over m^2_{\sc_R}} \right) +
  {1 \over m^2_{\st_R} } \ \left( y_t^2  - {|T_{U32}|^2 \over m^2_{\sc_L}} \right) -  {|T_{U33}|^2 \over m^2_{\st_L}  m^2_{\st_R}}  \Bigg]  \, ,
 \label{DEV(g)_approx}
\end{equation}
where $y_t = \sqrt{2}\, m_t/v_2 = g\, m_t/(\sqrt{2}\, m_W \sin\beta)$ is the top-quark
Yukawa coupling, $v = \sqrt{v_1^2 + v_2^2} =  2\, m_W/g =$~ 242 GeV is the vacuum
expectation value, $m_t$ is the top-quark mass, $m_W$ is the W-boson mass,
and $g$ is the SU(2) gauge coupling constant.
In Eq.~(\ref{DEV(g)_approx}) we have neglected terms $\propto \mu/\tan\beta$ because we use 
in this numerical study large values of $\tan\beta$ ($\ge 10$), see Eq.~(\ref {M2sqdef}). 
Note that Eq.~(\ref {DEV(g)_approx}) is not a function of $M^2_{U23}$ and $M^2_{Q23}$.
The terms $m^2_{\sc_{L,R}}$ and $m^2_{\st_{L,R}}$ are diagonal entries of the mass matrix 
${\cal M}^2_{\tilde{q}}$, Eq.~(\ref {EqMassMatrix1}). For values much larger than $v$ we can approximate them
by $m^2_{\sc_L} \simeq M^2_{Q 22}$,  $m^2_{\sc_R} \simeq M^2_{U 22}$, $m^2_{\st_L} \simeq M^2_{Q 33}$, and 
$m^2_{\st_R} \simeq M^2_{U 33}$. \\
From Eq.~(\ref{DEV(g)_approx}) we see that $DEV(g)^{approx}$ depends only on the squared absolute  
values of $T_{U23}, T_{U32}$, and $T_{U33}$. When all these three parameters go to zero, $DEV(g)^{approx}$  is small
and positive. For large values of $|T_{U23}|, |T_{U32}|$, and $|T_{U33}|$ $DEV(g)^{approx}$ becomes large and negative. Furthermore, 
$DEV(g)^{approx}$ also grows 
when $m^2_{\sc_{L,R}}$ and/or $m^2_{\st_{L,R}}$ decrease.

\begin{figure*}[t!]
\centering
\subfigure[]{
   { \mbox{\hspace*{-1cm} \resizebox{7.5cm}{!}{\includegraphics{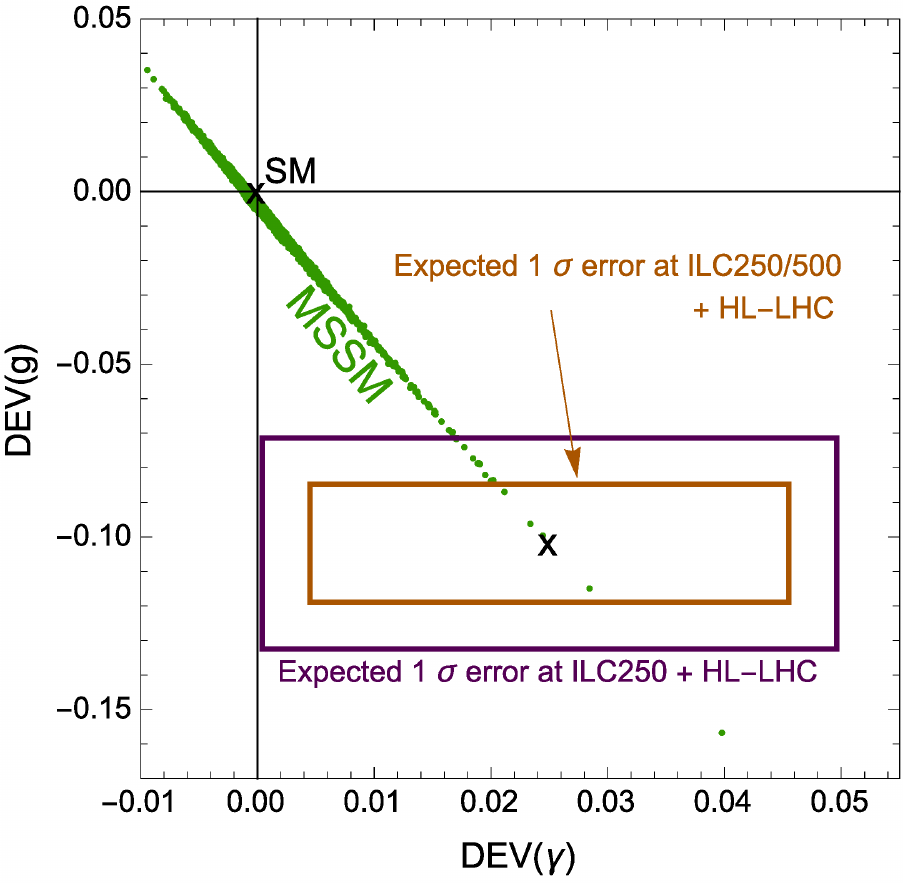}} \hspace*{0cm}}}
   \label{fig1a}}
 \subfigure[]{
   { \mbox{\hspace*{+0.cm} \resizebox{7.5cm}{!}{\includegraphics{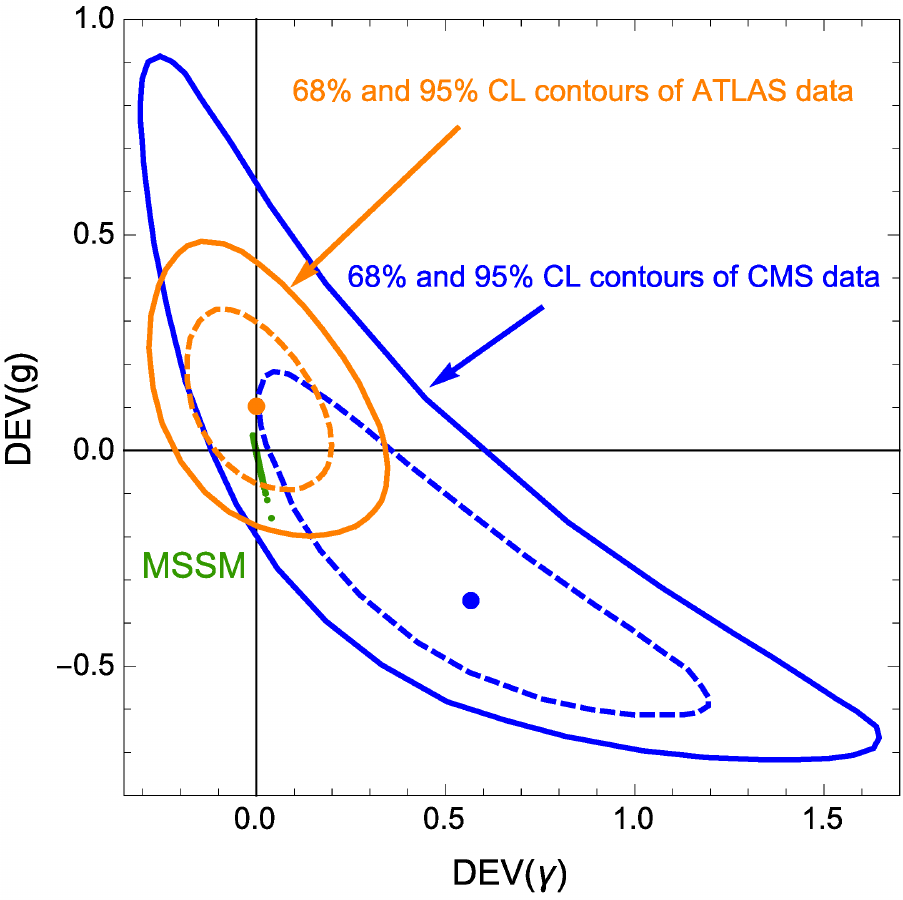}} \hspace*{-1cm}}}
  \label{fig1b}}\\
\caption{
The scatter plot of the scanned parameter points within the ranges given in Table \ref{table1} 
in the DEV($\gamma$) - DEV($g$) plane. 
(a): The expected 1$\sigma$ errors at ILC250/500 + HL-LHC [ILC250 + HL-LHC]; 
the black cross at (DEV($\gamma$), DEV($g$))=(0.025, -0.102) shows a possibly 
measured point of Eq.~(\ref{DEV_cdef}) and the orange and purple boxes indicate expected 1$\sigma$ errors of 
Eqs.~(\ref{DDEV_A}) and~(\ref{DDEV_B}) , respectively.
(b): The 68\% and 95\% CL contours of the recent ATLAS/CMS data\cite{ATLAS_kappa_plot,CMS_kappa_plot}.
}
\label{fig1}
\end{figure*}

In the following we show the results of a full parameter scan without
using this effective field theory approximation.

In Fig.~\ref{fig1} we show the scatter plot of the scanned parameter points within the ranges given in Table \ref{table1} 
in the $DEV(\gamma) - DEV(g)$ plane. 
We see that $DEV(g)$ is mostly negative and goes down to more than -10\%,
and that there is a strong correlation between $DEV(\gamma)$ 
and $DEV(g)$, 
\begin{equation}
DEV(\gamma)  \simeq - {1 \over 4} DEV(g) \, .
\end{equation}
Thus we also have $DEV(\gamma)^{approx}  \simeq - {1 \over 4} DEV(g)^{approx}$.  
This feature is due to the fact that the amplitude for $h^0 \to  \gamma \gamma$ is dominated by the 
W-boson loop contribution. The second important contribution to $h^0 \to  \gamma \gamma$
stems from the top-quark loop. 
The decay $h^0 \to  g g$ is dominated by the top-quark loop contribution. 
In the scenarios we are interested in, the up-type squark loop 
contributions to $h^0 \to \gamma \gamma / g g$ can be large.
All other SUSY contributions are relatively small, giving 
together less than 0.5\% in our study. 
Hence both $DEV(\gamma)$ and $DEV(g)$ are dominated by the same common 
source (i.e. $\su_{1,2}$-loops) which together with the W-loop 
dominance leads to the strong correlation.

Qualitatively our results are consistent with $DEV(g)^{approx}$ and $DEV(\gamma)^{approx}$ but 
it is hard to compare directly numerically because of the different usage of the MSSM input 
parameters, see the description at the end of Section~\ref{sec:full scan}.\\
The large deviations shown in Fig.~\ref{fig1} can be experimentally observed at a 
future $e^+ e^-$ collider such as ILC~\cite{ILC250} and/or CLIC~\cite{CLICref}. 
The abbreviations "ILC250/500 + HL-LHC" and "ILC250 + HL-LHC" are explained below. 
In Fig.~\ref{fig1}(b) the recent LHC data of the coupling modifiers ($\kappa_\gamma$, $\kappa_g$) 
transformed into the $(DEV(\gamma), DEV(g))$ plane by using the relation 
$DEV(X) = \kappa_X^2 -1$ are shown, where $\kappa_X = C(h^0XX)/C(h^0XX)_{SM}$ 
with $C(h^0XX)$ being the coupling of $h^0XX$. 
It is seen that the errors of the LHC data are very large and both the SM and the MSSM are 
allowed by the ATLAS/CMS data on the $h^0$ couplings $C(h^0\gamma\gamma)$ and $C(h^0gg)$.\\
If the measured point at ILC + HL-LHC was around ($DEV(\gamma)$, $DEV(g)$) = (0.025, -0.10) 
as shown in Fig.~\ref{fig1}(a), then the data would disfavour the SM  and favour 
the MSSM. If the measured point was around ($DEV(\gamma)$, $DEV(g)$) = (-0.05, -0.10), 
then we could say that the data disfavours both the SM and the MSSM.

\begin{figure*}[h!]
\centering
\subfigure[]{
      { \mbox{\hspace*{0cm} \resizebox{7.5cm}{!}{\includegraphics{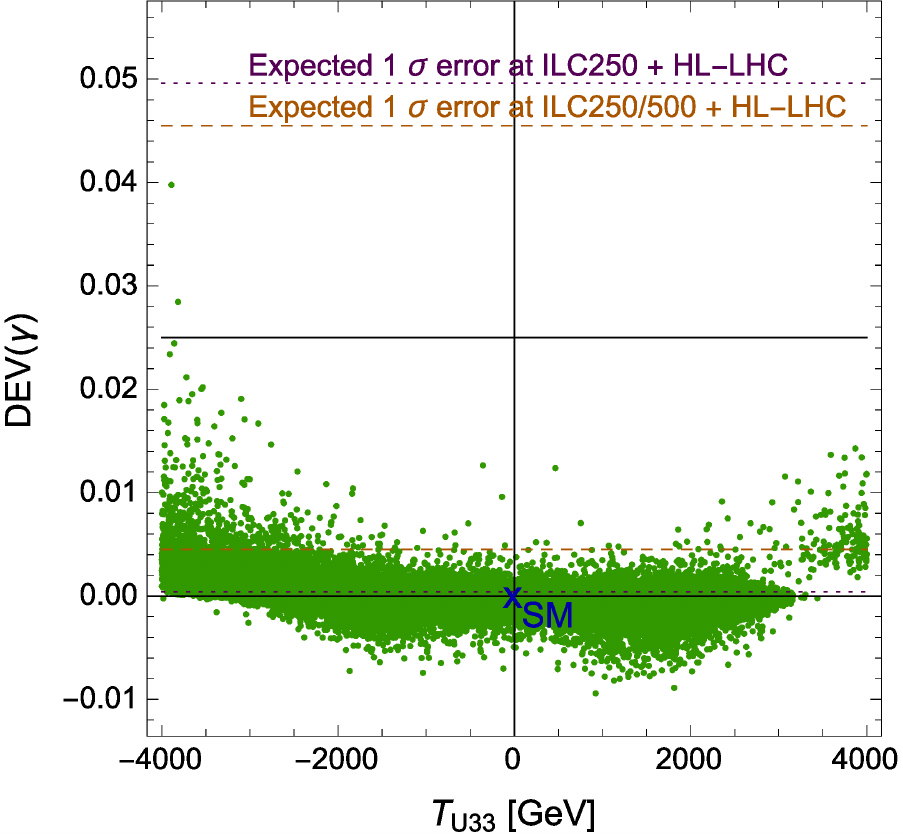}} \hspace*{-0.5cm}}}
   \label{fig2a}}  \hfill
 \subfigure[]{
   { \mbox{\hspace*{0cm} \resizebox{7.5cm}{!}{\includegraphics{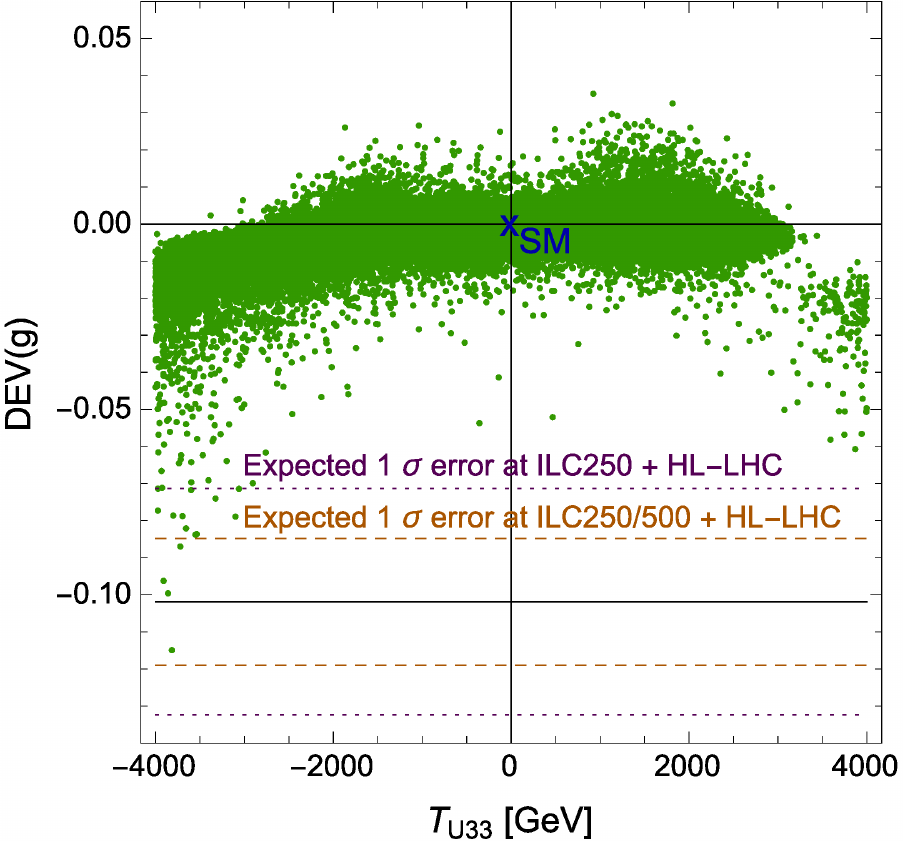}} \hspace*{-0.5cm}}}
   \label{fig2b}}\\
 \subfigure[]{
  { \mbox{\hspace*{0.cm} \resizebox{7.5cm}{!}{\includegraphics{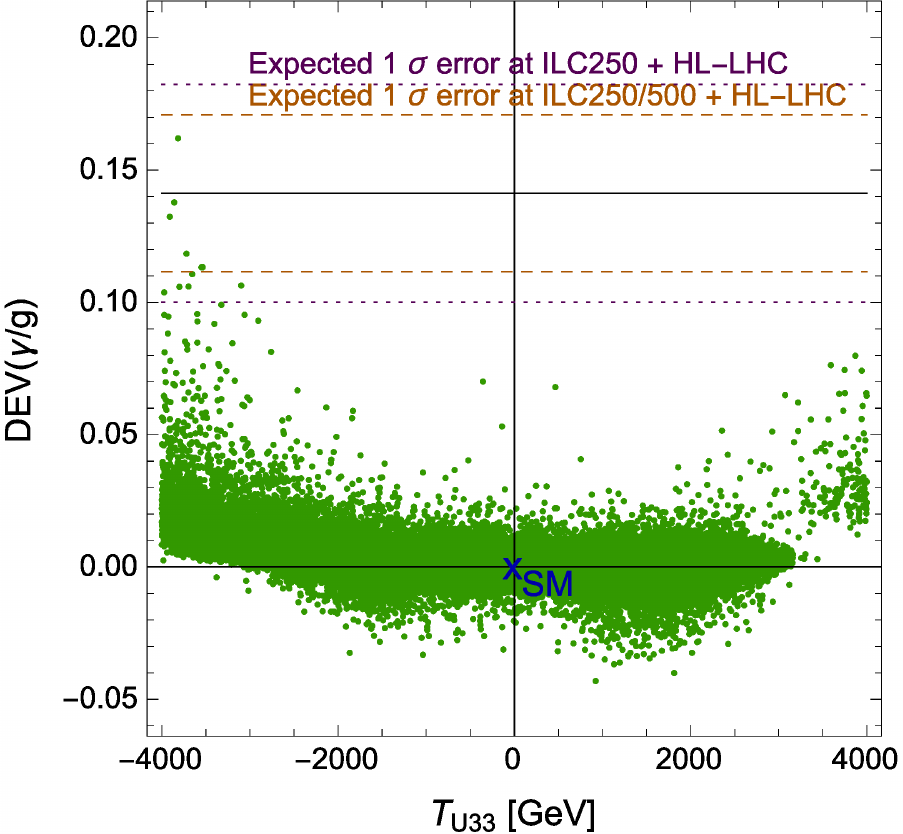}} \hspace*{0cm}}}
 \label{fig2c}}\\
\caption{
The scatter plots of the scanned parameter points within the ranges given in Table \ref{table1} in 
(a): $T_{U 33}$ - DEV($\gamma$); (b): $T_{U 33}$ - DEV($g$);
(c): $T_{U 33}$ - DEV($\gamma/g$) planes. The expected 1$\sigma$ errors at ILC250/500 + HL-LHC [ILC250 + HL-LHC]
are also shown.
The black horizontal solid lines at (DEV($\gamma$), DEV($g$), DEV($\gamma$/$g$))=(0.025, -0.102, 0.141) 
show possibly measured values of Eq.~(\ref{DEV_cdef}) and the orange and purple dashed-lines indicate expected 1$\sigma$ 
errors of Eqs.~(\ref{DDEV_A}) and~(\ref{DDEV_B}), respectively.}
\label{fig2}
\end{figure*}
In Fig.~\ref{fig2}  we show the scatter plots of the scanned parameter points within the ranges given 
in Table~\ref{table1} in the $T_{U 33}$ - DEV($\gamma$) (a), $T_{U 33}$ - DEV($g$) 
(b), and $T_{U 33}$ - DEV($\gamma/g$) (c) planes.  We see that DEV($g$) 
and DEV($\gamma/g$) can be large in the scanned parameter ranges for large values of$|T_{U 33}|$.
This means that the $\su_{1,2}$-loop ($\sim$ stop/scharm loops) contributions to these loop-induced 
decays are quite important.  As in Figure~\ref{fig1} the deviations shown can be observed at a 
future $e^+ e^-$ collider (ILC/CLIC).\\
In all three plots of Fig.~\ref{fig2} we see the parabolic increase of the $DEV$'s 
for increasing $|T_{U33}|$ as this is discussed after Eq.~(\ref{DEV(g)_approx}). 
The less populated region around $T_{U33} = 3$~TeV stems from the fact that the 
upper limit of the $m_{h^0}$ constraint is often violated there.

\begin{figure*}[h!]
\centering
\subfigure[]{
      { \mbox{\hspace*{0cm} \resizebox{7.5cm}{!}{\includegraphics{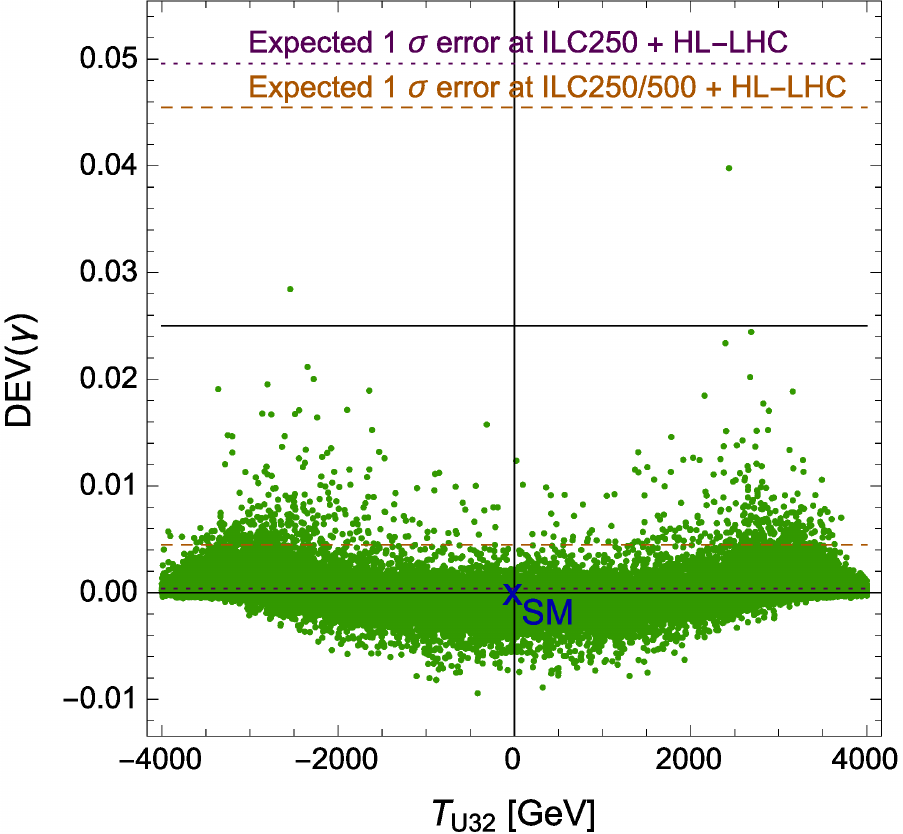}} \hspace*{-0.5cm}}}
   \label{fig3a}}  \hfill
 \subfigure[]{
   { \mbox{\hspace*{0cm} \resizebox{7.5cm}{!}{\includegraphics{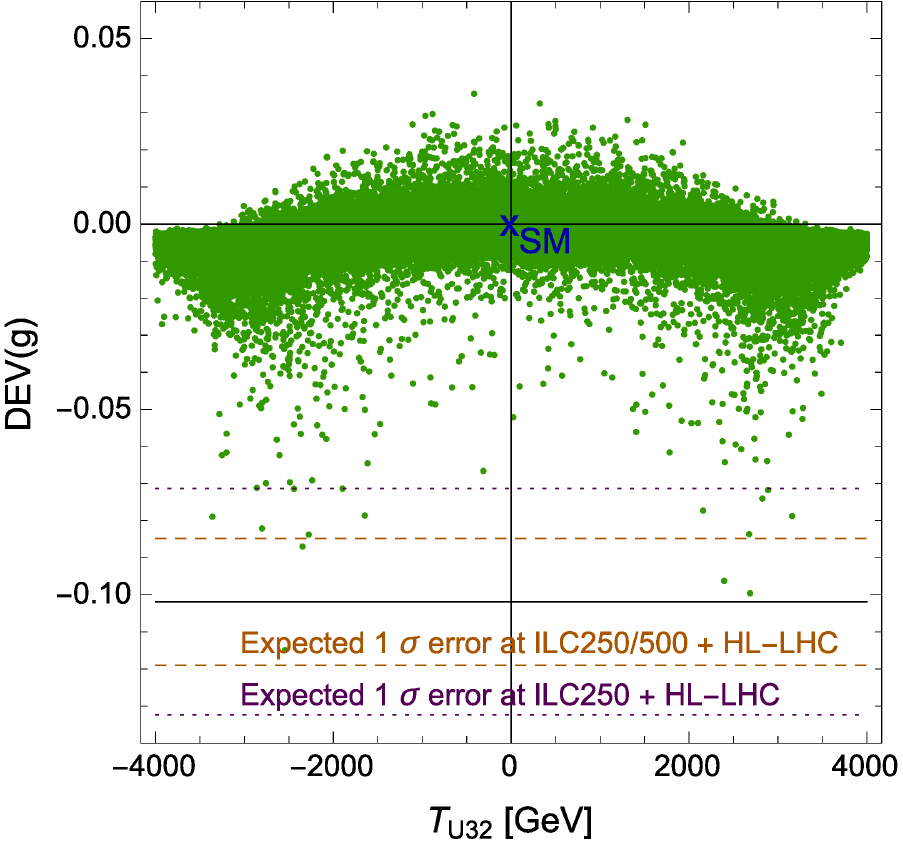}} \hspace*{-0.5cm}}}
   \label{fig3b}}\\
 \subfigure[]{
   { \mbox{\hspace*{0.cm} \resizebox{7.5cm}{!}{\includegraphics{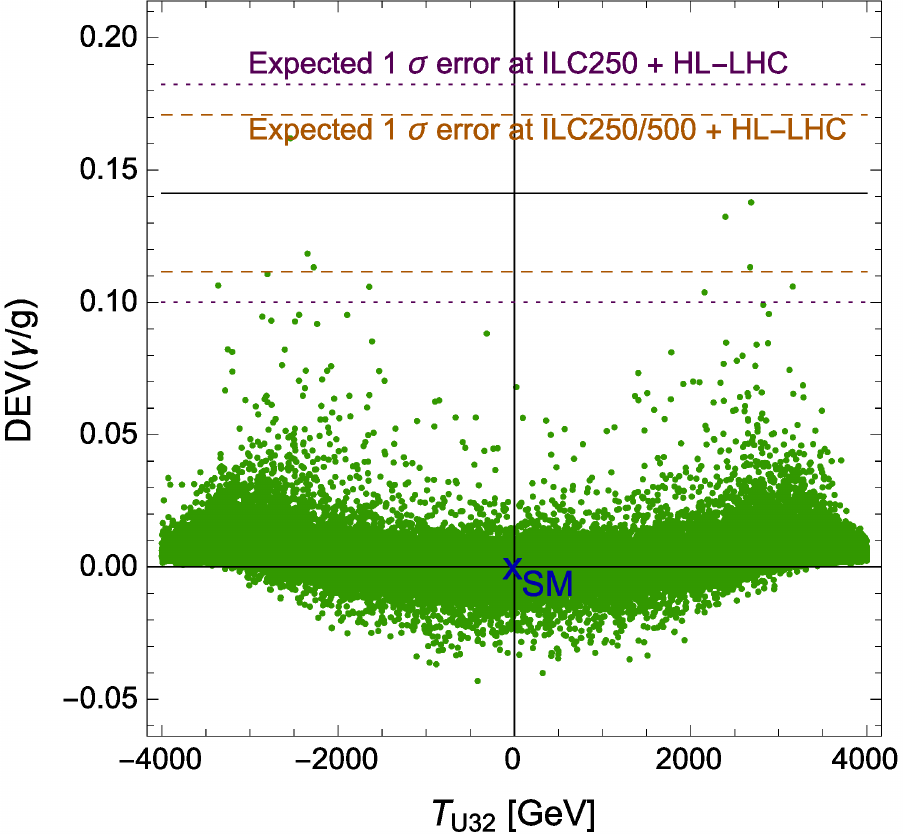}} \hspace*{0cm}}}
  \label{fig3c}}\\
\caption{
The scatter plot in the 
$T_{U 32}$ - DEV($\gamma$) (a), $T_{U 32}$ - DEV($g$) (b), and 
$T_{U 32}$ - DEV($\gamma/g$) (c) planes. 
The expected 1$\sigma$ errors at ILC250/500 + HL-LHC [ILC250 + HL-LHC]
are also shown as in Fig.~\ref{fig2}.
}
\label{fig3}
\end{figure*}

In order to show the importance of the QFV effect, in Fig.~\ref{fig3}
we show the scatter plot in the $T_{U 32} - DEV(\gamma)$~(a), 
$T_{U 32} - DEV(g)$~(b), and $T_{U 32} - DEV(\gamma/g)$~(c) planes.
In Fig.~\ref{fig3} we have a similar pattern as before in Fig.~\ref{fig2} 
but with the maximal results at slightly smaller values of the 
dependent variable, $|T_{U32}| \sim 2.5$~TeV. Again the parabolic shape is seen. 
And we see that in order to have large results we need the absolute value of both, 
the QFC parameter $T_{U33}$ and the QFV parameter $T_{U32}$, large.

We have obtained a similar dependence on $T_{U23}$ to that on $T_{U32}$.
Hence we do not show here the analogous plots on $T_{U23}$. 
In the parameter scan the average value of $M^2_{U22}$ is 1.8 times larger 
than that of $M^2_{U33}$. Therefore, the prefactor of $|T_{U23}|^2$ in Eq.~(\ref{DEV(g)_approx}) 
is 1.8 times smaller in average than that of $|T_{U32}|^2$ leading to somewhat milder $|T_{U23}|^2$ 
dependence of the deviations than that of $|T_{U32}|^2$. However, this choice of different mass 
ranges is just for a good efficiency (a good survival probability) of parameter scan in search 
for large deviations. The deviations can be enhanced by relatively light stop/scharm masses (see Eq.~(\ref{DEV(g)_approx})). 
Hence, relatively light mass ranges are taken for $M^2_{U22}$ and $M^2_{U33}$ in Table~\ref{table1}. 
In order to confirm that this choice does not affect our final conclusion essentially, we have 
performed the same parameter scan by taking common mass ranges [(0.6 TeV)$^2$, (4.0 TeV)$^2$] 
for \{$M^2_{Q22}, M^2_{Q33}, M^2_{U22}, M^2_{U33}, M^2_{D22}, M^2_{D33}$\}. We have obtained 
very similar scan results with slightly enhanced $T_{U23}$ dependence and much
smaller survival probability of the scan.

The common feature of the scan results is that the DEV's are 
significantly enhanced by the large values of the trilinear 
couplings $T_{U23}, T_{U32}, T_{U33}$.
This can be explained as follows:
\begin{itemize}

\item
The $ \sc_{R/L} - \st_{R/L}$ mixings can be large for large QFV parameters
$M^2_{Q 23}, M^2_{U 23}, T_{U 23}$, and $T_{U 32}$, for which the lighter up-type squarks
$\su_{1,2}$ can be strong mixtures of  $ \sc_{R/L} - \st_{R/L}$.

\item
In our decoupling Higgs scenario (with large $m_A$ and large $\tan\beta$), 
$h^0 \simeq  {\rm Re}(H_2^0)$ and hence $(T_{U 23},  T_{U 32},  T_{U 33})  \simeq (h^0 \st_L \sc_R, h^0 \st_R \sc_L, h^0 \st_L \st_R)$ couplings.

\end{itemize}

Thus, the $h^0 \su_{1,2} \su_{1,2}$ couplings and therefore also the $\su_{1,2}$-loop
contributions to $\Gamma(h^0 \to \gamma \gamma, g g)$ can be enhanced by
large $T_{U 23},  T_{U 32},  T_{U 33}$, which results in the significant
correlations between $T_{U 23},  T_{U 32},  T_{U 33}$, and $DEV(\gamma), DEV(g),
DEV(\gamma/g$). This explains the appearance of these $T_U$'s in Eq.~(\ref{DEV(g)_approx}).

Our analysis has shown that the correlations between the deviations DEV($\gamma$), DEV($g$), 
DEV($\gamma/g$) and all the FV/FC parameters other than those from the $\su$ sector, such as 
$T_{U23},T_{U32},T_{U33}$ and the stop/scharm masses, are very weak (see Eq.(\ref{DEV(g)_approx})). 
This means that the deviations DEV($\gamma$), DEV($g$), DEV($\gamma/g$) are quite 
insensitive to the parameters other than those of the up-type squark sector.
The latter is due to the fact that in our decoupling Higgs scenario $h^0 \simeq {\rm Re}(H_2^0)$.
Hence, the contributions of the down-type 
squark loops and the charged slepton loops to the decay widths $\Gamma(h^0 \to \gamma \gamma)$ 
and $\Gamma(h^0 \to g g)$ are very small. 
Note that $H_2^0$ couples to  $\st_L / \sca_L$ - $\st_R / \sca_R$ but does not to 
$\sb_L / \ti{s}_L$ - $\sb_R / \ti{s}_R$. 
Furthermore, for $DEV(\gamma)$, the charged Higgs 
and the chargino contributions always remain in the few-per mille range.

It is important to discuss the expected experimental errors.
We use two supposed data sets,
\mbox{data set A: ILC250/500 + HL-LHC}
and for collecting data without having a 500 GeV ILC,
\mbox{data set B: ILC250 + HL-LHC}.
The explanation of what "ILC250 + HL-LHC" and "ILC250/500 + HL-LHC" stand for is 
given in detail in the caption of Table~\ref{table1} of \cite{ILC250}, named there "ILC250" and "ILC500".
In order to discuss the experimental and theoretical errors we fix a possibly measured point,
\begin{equation}
\{DEV(\gamma)_c, DEV(g)_c, DEV(\gamma/g)_c\}  = \{ 2.5\%, -10.2\%, 14.1\%\} \, .
\label{DEV_cdef}
\end{equation}
This point is shown in 
Fig.~\ref{fig1}(a) by a black cross and in the Figs.~\ref{fig2}-\ref{fig3} by solid horizontal lines.
We use the relative estimated experimental 1$\sigma$~errors on the couplings $h \g\g$ and $h g g$ and their ratio
in the EFT fit framework,
\begin{eqnarray}
{\rm data set~A}: &&  \{ \d^r g_\g, \d^r g_g, \d^r g_{\g/g} \} = \{1\%, 0.95\%, 1.3\%\} \, ,\\
{\rm data set~B}: && \{ \d^r g_\g, \d^r g_g, \d^r g_{\g/g}\} = \{1.2\%, 1.7\%, 1.8\%\}\, ,
\end{eqnarray}
where $\delta^r y$ is defined as the relative error $\Delta y/y$ of the parameter $y$.
The values for $\d^r g_\g$ and $\d^r g_g$ are taken from Table~\ref{table1} in \cite{ILC250} and
the value for $\d^r g_{\g/g}$ we got from \cite{dga/g_error} using the same EFT fit program as in \cite{ILC250}.
Using
\begin{equation}
\Delta DEV(X) = 2 (DEV(X)_c + 1) \d^r g_X\, , \quad X = \g, g, \g/g\, ,
\end{equation}
we get the 1$\sigma$ errors for our $DEV$'s,
\begin{eqnarray}
{\rm data set~A}: && \{ \Delta DEV(\g), \Delta DEV(g), \Delta DEV(\g/g) \} = \{2.1\%, 1.7\%, 3.0\%\}\, ,
\label{DDEV_A}\\
{\rm data set~B}: && \{ \Delta DEV(\g), \Delta DEV(g), \Delta DEV(\g/g) \} = \{2.5\%, 3.1\%, 4.1\%\} 
\label{DDEV_B}\, .
\end{eqnarray}

\noindent
The 1$\sigma$ error bands are $DEV(X)_c \pm \Delta DEV(X)$, shown by boxes in Fig~\ref{fig1}a  and by 
dashed and dotted lines in Fig.~\ref{fig2} and Fig.~\ref{fig3}.
 
In all three figures Figs.~\ref{fig1}-\ref{fig3} we see that there are only a few dozens of points
where we have really a large deviation from the SM expectation values. This is just a matter of statistics because we
perform a scan in a 22-dimensional parameter space. Thus we choose a reference scenario 
where we have large $DEV$'s and then variate the most interesting parameters around this point P1.
All MSSM input parameters for P1 are shown in Table~\ref{table2} giving the
$DEV$'s in Eq.~(\ref{DEV_cdef}).

This scenario P1 satisfies all present experimental and theoretical constraints, see Appendix A. 
The resulting physical masses of the particles are shown in Table~\ref{physmasses}. 
For the calculation of the masses and the mixing, as well as for the low-energy observables, especially 
those in the B and K meson sectors (see Table~\ref{TabConstraints}), we use the public code 
{\tt SPheno} v3.3.8~\cite{SPheno1, SPheno2}. 
For the calculation of the coupling modifier $\kappa_b = C(h^0 b \bar{b})/C(h^0 b \bar{b})_{SM}$ 
(or equivalently the deviation $DEV(b)(= \kappa_b^2 -1)$ of the width $\Gamma(h^0 \to b \bar{b})$ from its SM value)
we compute the width $\Gamma(h^0 \to b \bar{b})$ at full one-loop level in the MSSM with QFV 
by using the code developed by us \cite{Eberl:h2bb}.  
We obtain $\kappa_b = 0.927$ (or $DEV(b) = -0.141$) which 
satisfies the LHC data in Table~\ref{TabConstraints}. For the B and K meson observables we get;
$B(b \to s \gamma) = 3.177 \cdot 10^{-4}$, $B(b \to s \ l^+ l^-) = 1.588 \cdot 10^{-6}$, 
$B(B_s \to \mu^+ \mu^-) = 3.065 \cdot 10^{-9}$, $B(B^+ \to \tau^+ \nu) = 9.956 \cdot 10^{-5}$, 
$\Delta M_{B_s} = 19.606 [ps^{-1}]$, $|\epsilon_K| = 2.205 \cdot 10^{-3}$, $\Delta M_K = 2.322 \cdot 10^{-15} \ (GeV)$, 
$B(K^0_L \to \pi^0 \nu \bar{\nu}) = 2.307 \cdot 10^{-11}$, and $B(K^+ \to \pi^+ \nu \bar{\nu}) = 7.734 \cdot 10^{-11}$, 
all of which satisfy the constraints of Table~\ref{TabConstraints}.

%
\begin{table}[h!]
\footnotesize{
\caption{The MSSM parameters for the reference point P1 (in units 
of GeV or GeV$^2$ expect for $\tan\beta$)}
\begin{center}
\begin{tabular}{|c|c|c|c|c|c|}
    \hline
\vspace*{-0.3cm}
& & & & &\\
\vspace*{-0.3cm}
     $\tan\beta$ & $M_1$ &  $M_2$ & $M_3$ & $\mu$ &  $m_A(pole)$\\ 
& & & & &\\
    \hline
\vspace*{-0.3cm}
& & & & &\\
\vspace*{-0.3cm}
     16 & 1270 & 500  & 4800 & 1260 & 1960\\
& & & & &\\
    \hline
    \hline
\vspace*{-0.3cm}
& & & & &\\
\vspace*{-0.3cm}
      $M^2_{Q 22}$ & $ M^2_{Q 33}$ &  $M^2_{Q 23}$ & $ M^2_{U 22} $ & $ M^2_{U 33} $ &  $M^2_{U 23} $\\ 
& & & & &\\
     \hline
\vspace*{-0.3cm}
& & & & &\\
\vspace*{-0.3cm}
     3660$^2$ &  2520$^2$ & 550$^2$ & 3710$^2$ & 1435$^2$ & 875$^2$\\
& & & & &\\
    \hline
    \hline
\vspace*{-0.3cm}    
& & & & &\\
\vspace*{-0.3cm}      
      $ M^2_{D 22} $ & $ M^2_{D 33}$ &  $ M^2_{D 23}$ & $T_{U 23}  $ & $T_{U 32}  $ &  $T_{U 33}$\\ 
& & & & &\\
    \hline
\vspace*{-0.3cm}      
& & & & &\\
\vspace*{-0.3cm}  
      3620$^2$ & 2720$^2$ & 925$^2$ & 760 & 1560 & - 4200\\
& & & & &\\
 \hline 
\multicolumn{6}{c}{}\\[-3.6mm]  
\cline{1-4}
\vspace*{-0.3cm}      
     & & & \\
\vspace*{-0.3cm}      
     $ T_{D 23} $ & $T_{D 32}  $ &  $ T_{D 33}$ &$T_{E 33} $\\ 
     & & & \\
    \cline{1-4}
\vspace*{-0.3cm}      
     & & & \\
\vspace*{-0.3cm}      
     -565 & 690 & 270 & - 470\\
     & & & \\
    \cline{1-4}
\end{tabular}\\[3mm]
\begin{tabular}{|c|c|c|c|c|c|c|c|c|}
    \hline
\vspace*{-0.3cm}      
    & & & & & & & &\\
\vspace*{-0.3cm}      
    $M^2_{Q 11}$ & $M^2_{U 11} $ &  $M^2_{D 11} $ & $M^2_{L 11}$ & $M^2_{L 22} $ &  $M^2_{L 33}$ & $M^2_{E 11}$&$M^2_{E 22}$ & $M^2_{E 33} $\\ 
    & & & & & & & &\\
    \hline
\vspace*{-0.3cm}      
    & & & & & & & &\\
\vspace*{-0.3cm}      
    $4500^2$ & $4500^2$ & $4500^2$  & $1500^2$ & $1500^2$ & $1500^2$& $1500^2$& $1500^2$&$1500^2$\\
    & & & & & & & &\\
    \hline
\end{tabular}
\end{center}
\label{table2}
}
\end{table}

\begin{table}
\caption{Physical masses in GeV of the particles for the scenario of Table~\ref{table2}.}
\begin{center}
\begin{tabular}{|c|c|c|c|c|c|}
  \hline
  $\mnt{1}$ & $\mnt{2}$ & $\mnt{3}$ & $\mnt{4}$ & $\mch{1}$ & $\mch{2}$ \\
  \hline \hline
  $532.1$ & $1242$ & $1271$ & $1310$ & $532.3$ & $1275$ \\
  \hline
\end{tabular}
\vskip 0.4cm
\begin{tabular}{|c|c|c|c|c|}
  \hline
  $m_{h^0}$ & $m_{H^0}$ & $m_{A^0}$ & $m_{H^+}$ \\
  \hline \hline
  $125.5$  & $1960$ & $1960$ & $1962$ \\
  \hline
\end{tabular}
\vskip 0.4cm
\begin{tabular}{|c|c|c|c|c|c|c|}
  \hline
  $\msg$ & $\msu{1}$ & $\msu{2}$ & $\msu{3}$ & $\msu{4}$ & $\msu{5}$ & $\msu{6}$ \\
  \hline \hline
  $4562$ & $725$ & $2204$ & $3497$ & $3551$ & $4380$ & $4386$ \\
  \hline
\end{tabular}
\vskip 0.4cm
\begin{tabular}{|c|c|c|c|c|c|}
  \hline
 $\msd{1}$ & $\msd{2}$ & $\msd{3}$ & $\msd{4}$ & $\msd{5}$ & $\msd{6}$ \\
  \hline \hline
  $2173$ & $2421$ & $3467$ & $3497$ & $4380$ & $4386$ \\
  \hline
\end{tabular}
\end{center}
\label{physmasses}
\end{table}
%

\begin{figure*}[h!]
\centering
\subfigure[]{
   { \mbox{\hspace*{-0.5cm} \resizebox{7.5cm}{!}{\includegraphics{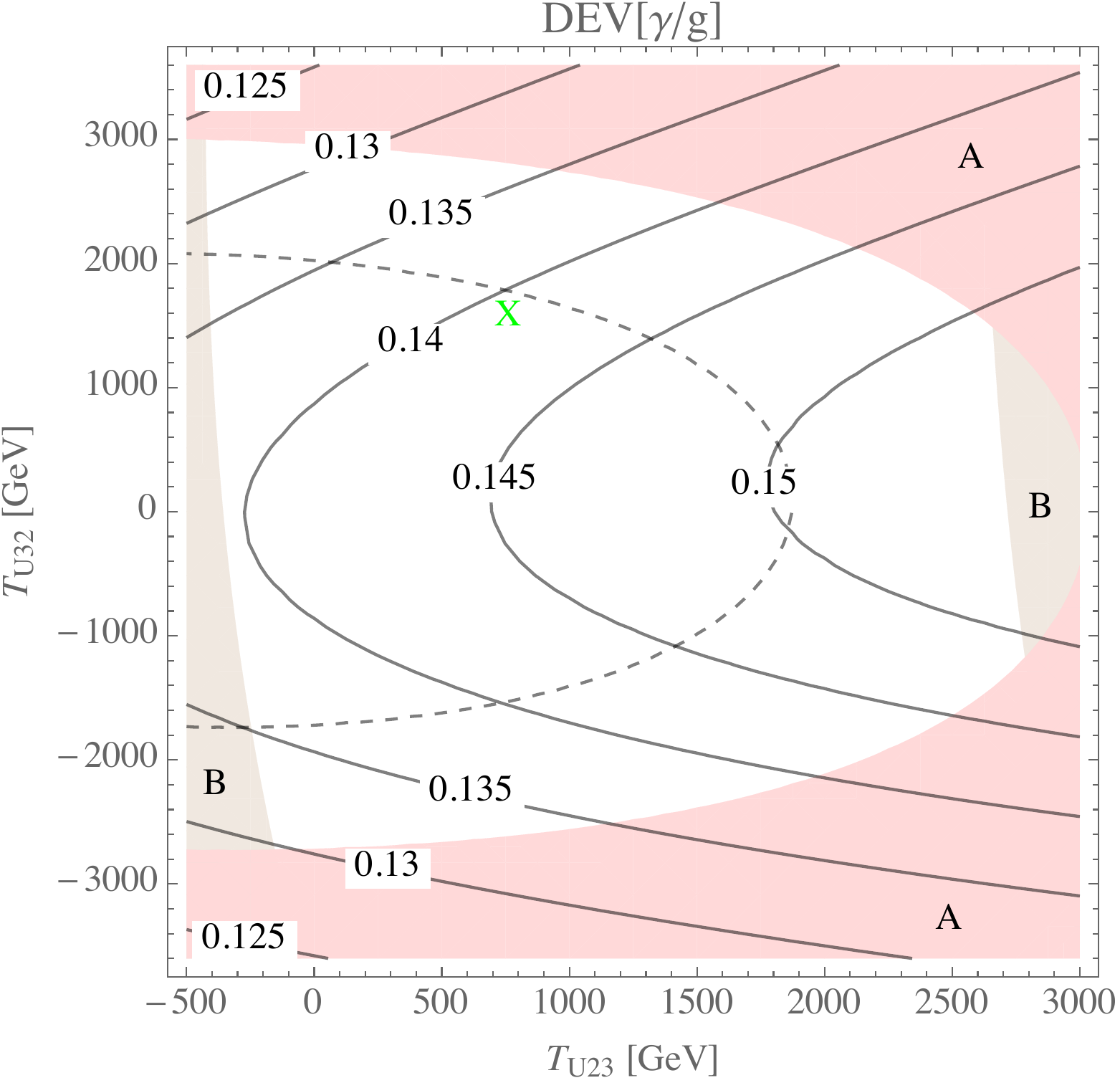}} \hspace*{0cm}}}
   \label{fig4a}} 
 \subfigure[]{
   { \mbox{\hspace*{0cm} \resizebox{7.1cm}{!}{\includegraphics{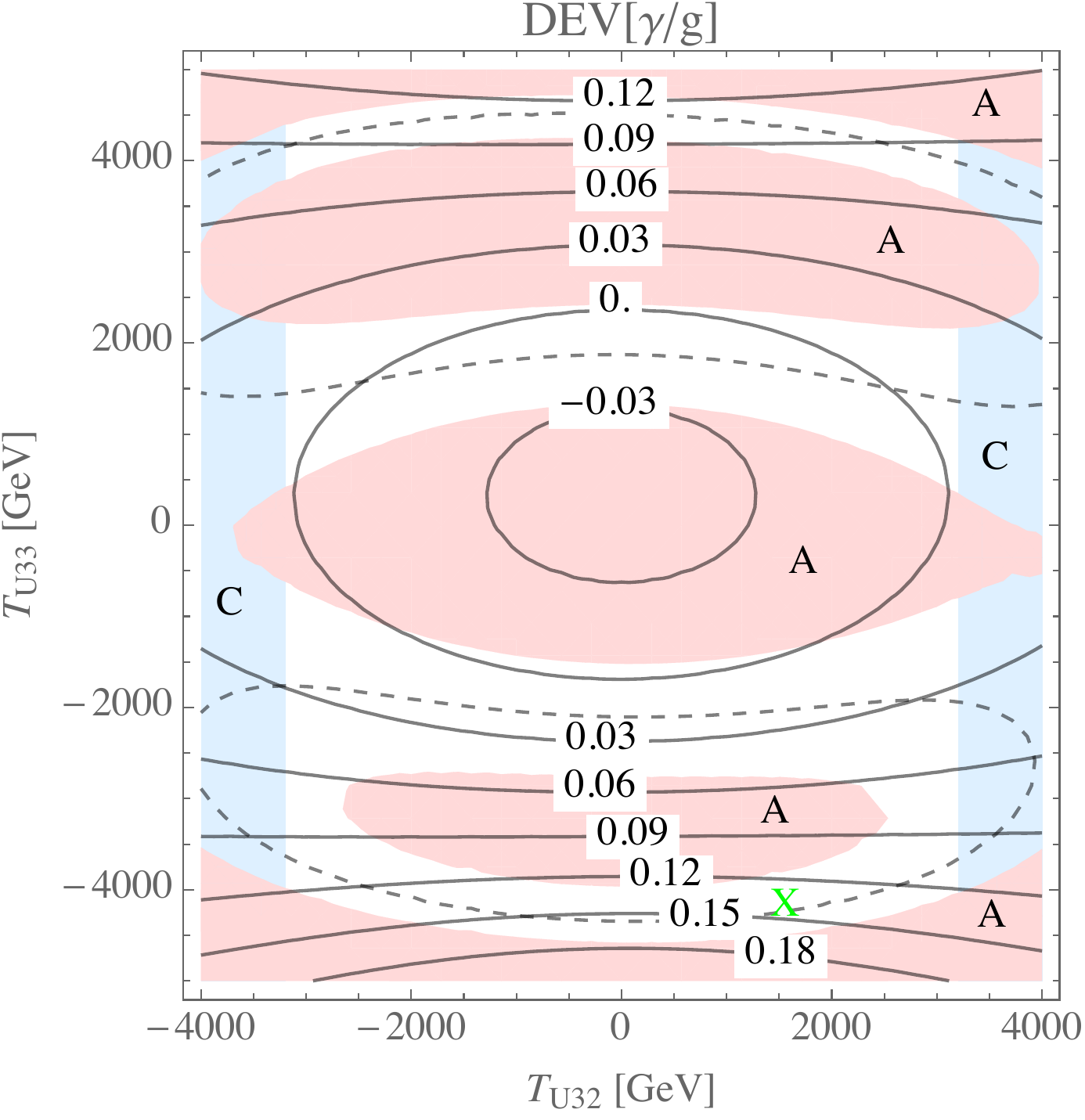}} \hspace*{-0.5cm}}}
   \label{fig4b}}\\
 \subfigure[]{
   { \mbox{\hspace*{0cm} \resizebox{7.5cm}{!}{\includegraphics{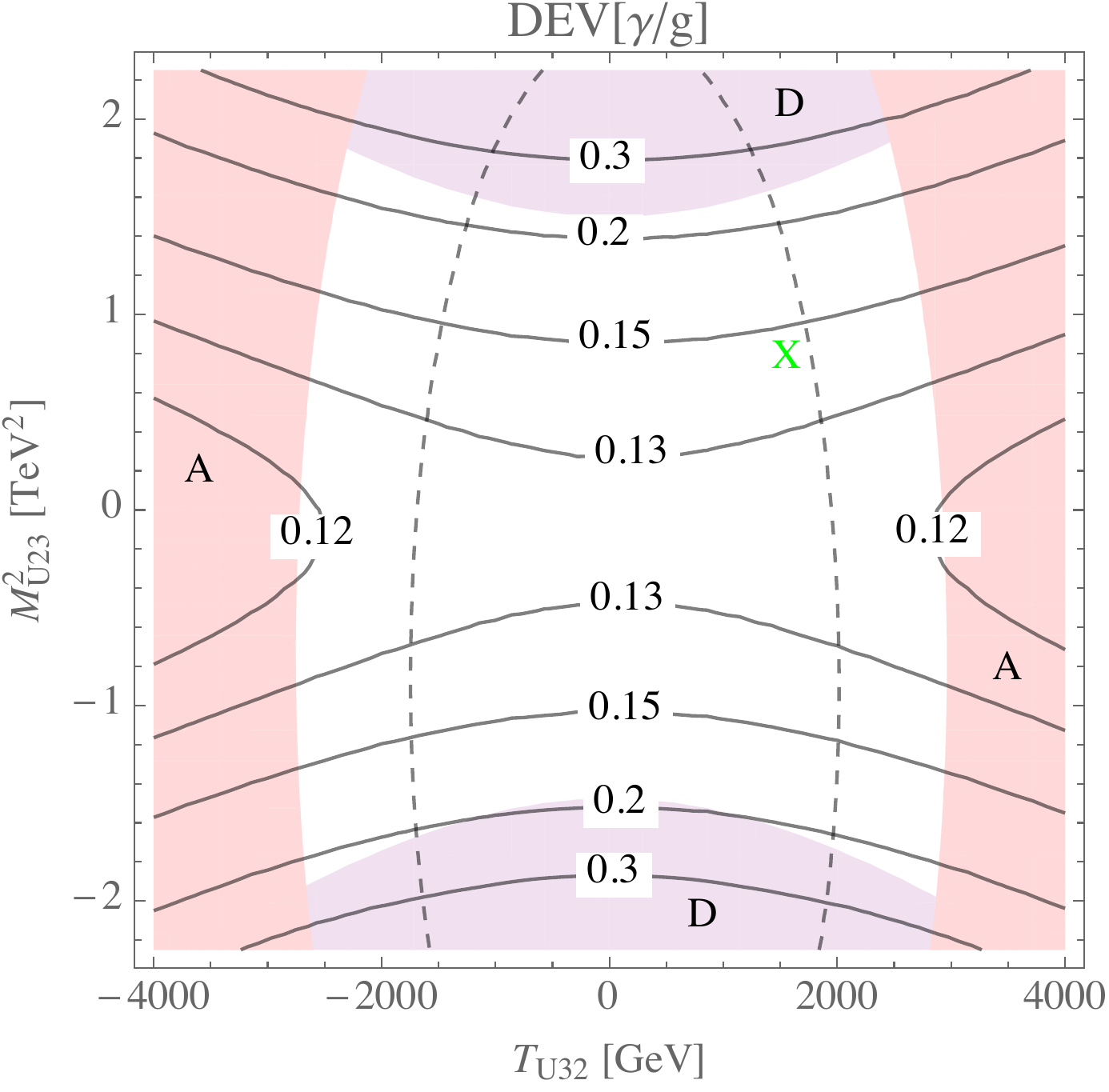}} \hspace*{0cm}}}
  \label{fig4c}}\\
\caption{Contour plots of DEV($\gamma/g$) in the 
     $T_{U 32}$ - $T_{U 23}$ (a), $T_{U 32}$ - $T_{U 33}$ (b), $T_{U 32}$ - $M^2_{U 23}$ (c) planes. 
     The parameters other than the shown ones in each plane are fixed as in Table~\ref{table2}.
     The "X" marks P1 in the plots.
     The shown forbidden areas are due to the constraints: $A \equiv m_{h^0}$, $B \equiv {\rm B}(B_s \to \mu^+ \mu^-)$, 
     $C \equiv$ vacuum stability condition, $D \equiv m_{\su_1}$. 
     The dashed lines are the contours of $m_{h^0} = 125.09$~GeV.
     }
\label{fig4}
\end{figure*}

In Fig.~\ref{fig4} we show the contour plots of DEV($\gamma/g$) in the QFV/QFC parameter plane around P1. 
The reference point is marked by a green "x". We see that $DEV(\gamma/g)$ is really large in a
large region of the parameter planes and that the effect of
the QFV parameters $M^2_{U23}, T_{U23}, T_{U32}$ (and the QFC parameter
$T_{U 33}$ also) on the $DEV(\gamma/g)$ is very important.
We again see the parabolic behaviour on all the $T_U$ parameters. For this parameter point the dependence on 
$T_{U 32}$ and $T_{U 23}$ is of similar size and the dependence on $T_{U 33}$ varies from 
-3\% up to 16\% in the allowed region. Fig.~\ref{fig4}(c) shows a strong dependence on $\sc_R-\st_R$ mixing parameter 
$M^2_{U23}$ which means that for large $M^2_{U23}$ the "linearized" approximation Eq.~(\ref{DEV(g)_approx}) is not good anymore. 
There one should add higher orders to Eq.~(\ref{DEV(g)_approx}) which includes $M^2_{U23}$.

Finally, we also discuss the theoretical errors.
The theoretical uncertainties of the MSSM predictions are twofold.
If we consider a fixed MSSM parameter point, the 
total theoretical error can be split into two parts: one is the 
uncertainty due to unknown (higher order) loop contributions and 
the other one - the uncertainty due to errors of the SM input parameters. 
The former uncertainty we call scale uncertainty and the latter 
one - parametric uncertainty. 
The scale uncertainty can be estimated 
by varying the renormalization scale $Q$ from $Q = m_{h^0}/2$ up to $Q = 2 m_{h^0}$.

We can write the relative parametric uncertainty as
\begin{equation}
\delta^{r,P} DEV(X) =  \bigg|{ m_t \over DEV(X)} {\partial DEV(X)  \over \partial m_t}\bigg|\delta^r m_t \oplus
\bigg|{ \alpha_s \over DEV(X)} {\partial DEV(X)  \over \partial \alpha_s}\bigg| \delta^r \alpha_s \, ,
\end{equation}
with $X = \g, g, \g/g$. 
We have found that we can neglect the parametric uncertainties due to all
the other SM parameters such as  $m_b$, $\alpha_{EM}$, $m_Z$ etc..
We use as input the on-shell top-mass, $m_t = 173$~GeV with
$\d^r m_t = 0.23$\%, and $\alpha_s \equiv \alpha_s(m_Z)_{\msbar} = 0.1181$ with 
$\d^r \alpha_s = 0.93$\% \cite{PDG2018}. We get for the reference point P1 at $1\sigma$
\begin{eqnarray*}
\delta^{r,P} DEV(\g)  & = & |- 1.7| \d^r m_t  \oplus  | 3.0| \d^r \a_s = \hphantom{0}0.4\% \oplus 2.8\%\, ,\\
\delta^{r,P} DEV(g)  & = & |- 0.2| \d^r m_t  \oplus  | 2.8| \d^r \a_s = 0.05\% \oplus 2.6\%\, ,\\
\delta^{r,P} DEV(\g/g)  & = & |-0.5| \d^r m_t  \oplus  | 3.1| \d^r \a_s = \hphantom{0}0.1\% \oplus 2.9\%\, .
\end{eqnarray*}
One would guess that for $DEV(\g)$ there should be a small coefficient in front of $\d^r \a_s$. 
This is not the case because $\a_s$ has a strong influence on the calculation of the 
running top Yukawa coupling at $Q = m_{h^0}$ and on that of the $\su$~parameters entering 
the $h^0 \su \su^*$~couplings.\\
From the scale variation we get 
\begin{eqnarray*}
\delta^{r,Q} DEV(\g)  & = & \begin{array}{c}  \hphantom{-}2.3 \% \\ -2.1\% \end{array}  \simeq 2.3\%\, ,\\
\delta^{r,Q} DEV(g)  & = &\begin{array}{c}   \hphantom{-}2.9\% \\ -2.6\% \end{array}  \simeq 2.9\%\, ,\\
\delta^{r,Q} DEV(\g/g)  & = & \begin{array}{c}  \hphantom{-}3.2\% \\ -2.8\% \end{array}  \simeq 3.2\%\, .
\end{eqnarray*}
The upper value is for $Q = m_{h^0}/2$ and the lower one for $Q = 2 m_{h^0}$.
Thus we estimate the total theoretical relative and absolute errors $\Delta DEV(X) = \d^r DEV(X) DEV(X)_c$,
at $1\sigma$ for the point P1,
\begin{eqnarray*}
\delta^{r} DEV(\g)  =  5.1\% \, , & \quad \Delta DEV(\g) &  = 0.13\%\,, \\
\delta^{r} DEV(g)   =  5.5\%  \, , & \quad \Delta DEV(g) &  = 0.55\%\,,\\
\delta^{r} DEV(\g/g)  = 6.1\% \, , & \!\!\!\!\! \quad \Delta DEV(\g/g) &  = 0.85\%\, ,
\end{eqnarray*}
where the parametric uncertainties are added quadratically and the scale uncertainty
is added to them linearly.
Comparing this result with Eqs.~(\ref{DDEV_A}) and (\ref{DDEV_B}) we see 
that the theoretical errors are one order smaller than the 
experimental ones at P1. From Eqs. (\ref{DEV_cdef}), (\ref{DDEV_A}), (\ref{DDEV_B}) 
and the theoretical errors, we see that ILC cannot miss this SUSY signal in case the scenario P1 
(or similar ones) is realized in Nature.

Using the LO (lowest order) results instead of the NLO results at P1, the
relative shifts of the DEV's are found to be very small (less than 1\%).
This is due to the fact that in our computation the NLO QCD corrections
are included only in the SM parts which dominate the MSSM widths.

One might think that the experimental and theoretical 
improvement expected in the low-energy observables could exclude
the flavour-violating squark scenarios in the first place, way 
before the beginning of the HL-LHC or the ILC. 
On the other hand, the low-energy observables have both 
experimental and theoretical errors and currently the latter 
errors tend to be comparable to (or larger than) the former ones
as shown in Table \ref{TabConstraints}. The theoretical improvement expected in the 
low-energy observables is rather unclear. 
Only if the observed values with almost zero errors perfectly agree 
with the SM predictions with almost zero errors, the possibility of the flavour-violating squark scenarios will be excluded.

\section{Conclusions}
\label{sec:concl}

We have studied the correlation between the loop-induced decays 
$h^0 \to \gamma \gamma$ and $h^0 \to g g$ in the MSSM with QFV. 
From a full parameter scan and a detailed analysis around a fixed reference point, 
respecting all the relevant theoretical and 
experimental constraints, we have found that
\begin{itemize}
\item the relative deviation of the MSSM decay width $\Gamma(h^0 \to g \, g)$ from 
the Standard Model value, $DEV(g)$, can be large and
negative down to $\sim$ -15\% in the studied parameter ranges,

\item there is a strong correlation between $DEV(\gamma)$ and $DEV(g)$,

\item the relative deviation of the width ratio $DEV(\gamma/g)$ from the
  SM value can be large (up to $\sim$ 20\%) in the studied parameter ranges,

\item both SUSY QFV and QFC up-type squark parameters can have a strong influence on 
these deviations and their contributions add up.

\end{itemize}

Such large deviations can be observed at a future $e^+ e^-$ collider such as ILC and CLIC. 
Observation of the deviation patterns as shown in this 
study would favour the MSSM with flavour-violating squark mixings 
and encourage to perform further studies in this model.

%
\section*{Acknowledgments}

We would like to thank W. Porod for helpful discussions, especially for the 
permanent support concerning SPheno. We also thank J.~Tian for sharing
his expertise on ILC physics with us.
We also thank Prof. A. Bartl for useful discussions at the 
early stage of this work.\\
VRVis is funded by BMVIT, BMDW, Styria, SFG and Vienna Business Agency in the 
scope of COMET - Competence Centers for Excellent Technologies (854174) which is managed by FFG.

\begin{appendix}
 
\section{Theoretical and experimental constraints}
\label{sec:constr}
The experimental and theoretical constraints taken into account in the 
present work are discussed in detail in~\cite{Eberl_17}. Here we only 
list the updated constraints from K- and B-physics and those on the Higgs 
boson mass and coupling in Table~\ref{TabConstraints}. 

The $h^0$ couplings that receive SUSY QFV effects significantly are $C(hbb)$ ~\cite{Eberl:h2bb}, $C(hcc)$ ~\cite{Bartl:2014bka}, 
$C(hgg)$ and $C(h\gamma\gamma)$ 
\footnote{
Precisely speaking, in principle, $C(htt)$ coupling could also 
receive SUSY QFV effects significantly. 
However, predicting the (effective) coupling $C(htt)$ at loop levels in the MSSM 
is very difficult since its theoretical definition in the context of tth production at LHC is unclear ~\cite{tth@LHC}.
}. The measurement of $C(hcc)$ is very difficult 
due to huge QCD backgrounds at LHC; there is no significant experimental data 
on $C(hcc)$ at this moment. Hence, the relevant h couplings to be compared 
with the LHC observations are $C(hbb)$, $C(hgg)$ and $C(h\gamma\gamma)$. 
The MSSM predictions for the couplings $C(hgg)$ and $C(h\gamma\gamma)$ are 
allowed by the current LHC data as shown in Fig.~\ref{fig1}(b). Therefore, we list 
the LHC data on $C(hbb)$ ($\kappa_b$) in Table~\ref{TabConstraints}.

In \cite{Dedes} the QFV decays $t \to q h$ with $q = u, c$, have 
been studied in the general MSSM with QFV. It is found that these decays cannot 
be visible at the current and high luminosity LHC runs due to the very small 
decay branching ratios B($t \to q h$).

\noindent In addition to these we also require our scenarios to be 
consistent with the following updated experimental constraints:

\begin{table*}[h!]
\footnotesize{
\caption{
Constraints on the MSSM parameters from the K- and B-meson data 
relevant mainly for the mixing between the second and the third generations of 
squarks and from the data on the $h^0$ mass and coupling $\kappa_b$. The fourth 
column shows constraints at $95 \%$ CL obtained by combining the experimental error 
quadratically with the theoretical uncertainty, except for $B(K^0_L \to \pi^0 \nu \bar{\nu})$, 
$m_{h^0}$ and $\kappa_b$.
}
\begin{center}
\begin{tabular}{|c|c|c|c|}
    \hline
    Observable & Exp.\ data & Theor.\ uncertainty & \ Constr.\ (95$\%$CL) \\
    \hline\hline
    &&&\\
    $10^3\times|\epsilon_K|$ & $2.228 \pm 0.011$ (68$\%$ CL)~\cite{PDG2019} 
    & $\pm 0.28$ (68$\%$ CL)~\cite{epsK_DMK_SM} &
    $2.228 \pm 0.549$\\
    $10^{15}\times\Delta M_K$ [GeV] & $3.484\pm 0.006$ (68$\%$ CL)~\cite{PDG2019} 
    & $\pm 1.2 $ (68$\%$ CL)~\cite{epsK_DMK_SM} &
    $3.484 \pm 2.352$\\
    $10^{9}\times$B($K^0_L \to \pi^0 \nu \bar{\nu}$) & $< 3.0$ (90$\%$ CL)~\cite{PDG2019} 
    & $\pm 0.002 $ (68$\%$ CL)~\cite{PDG2019} &
    $< 3.0$ (90$\%$ CL)\\
    $10^{10}\times$B($K^+ \to \pi^+ \nu \bar{\nu}$) & $1.7 \pm 1.1$ (68$\%$ CL)~\cite{PDG2019} 
    & $\pm 0.04 $ (68$\%$ CL)~\cite{PDG2019} &
    $1.7^{+2.16}_{-1.70}$\\
    $\Delta M_{B_s}$ [ps$^{-1}$] & $17.757 \pm 0.021$ (68$\%$ CL)~\cite{HFAG2016} 
    & $\pm 2.7$ (68$\%$ CL)~\cite{DeltaMBs_SM} &
    $17.757 \pm 5.29$\\
    $10^4\times$B($b \to s \gamma)$ & $3.49 \pm 0.19$ (68$\%$ CL)~\cite{HFAG2016, PDG2016} 
    & $\pm 0.23$ (68$\%$ CL)~\cite{Misiak_2015} &  $3.49\pm 0.58$\\
    $10^6\times$B($b \to s~l^+ l^-$)& $1.60 ~ ^{+0.48}_{-0.45}$ (68$\%$ CL)~\cite{bsll_BABAR_2014}
    & $\pm 0.11$ (68$\%$ CL)~\cite{Huber_2008} & $1.60 ~ ^{+0.97}_{-0.91}$\\
    $(l=e~{\rm or}~\mu)$ &&&\\
    $10^9\times$B($B_s\to \mu^+\mu^-$) & $2.8~^{+0.7}_{-0.6}$ (68$\%$CL)~\cite{Bsmumu_LHCb_CMS}
    & $\pm0.23$  (68$\%$ CL)~\cite{Bsmumu_SM_Bobeth_2014} 
    & $2.80~^{+1.44}_{-1.26}$ \\
    $10^4\times$B($B^+ \to \tau^+ \nu $) & $1.14 \pm 0.27$ (68$\%$CL)
    ~\cite{Trabelsi_EPS-HEP2015, Hamer_EPS-HEP2015}
    &$\pm0.29$  (68$\%$ CL)~\cite{Btotaunu_LP2013} & $1.14 \pm 0.78$\\
    $ m_{h^0}$ [GeV] & $125.09 \pm 0.24~(68\%~ \rm{CL})$ \cite{Higgs_mass_ATLAS_CMS}
    & $\pm 3$~\cite{Higgs_mass_Heinemeyer} & $125.09 \pm 3.48$ \\
    $\kappa_b$ & $1.06^{+0.37}_{-0.35}~(95\%~ \rm{CL})$ \cite{kappa_b_ATLAS}
    &  & $1.06^{+0.37}_{-0.35}$ (ATLAS)\\
    & $1.17^{+0.53}_{-0.61}~(95\%~ \rm{CL})$ \cite{kappa_b_CMS}
    &  & $1.17^{+0.53}_{-0.61}$ (CMS)\\
&&&\\
    \hline
\end{tabular}
\end{center}
\label{TabConstraints}}
\end{table*}
%
\begin{itemize}

\item
The LHC limits on sparticle masses (at 95\% CL)~\cite{SUSY@EPS-HEP2017}-\cite{Strandberg18}:

In the context of simplified models, gluino masses $\msg \lesssim 2.1~{\rm TeV}$ are 
excluded at 95\% CL. The mass limit varies in the range 1800-2100~GeV depending 
on assumptions. First and second generation squark masses are excluded below 1500~GeV. 
Bottom squark masses are excluded below 1250~GeV. A typical top-squark mass lower limit is 
$\sim$ 1100~GeV for $m_{\nt_1} < 500$ GeV. There is no top-squark mass limit for 
$m_{\nt_1} > 500$ GeV. 
For sleptons heavier than the lighter chargino $\ch_1$ and the second neutralino $\nt_2$, 
the mass limits are $m_{\ch_1}, m_{\nt_2} > 650$ GeV for $m_{\nt_1} \lesssim 300$ GeV and 
there is no $m_{\ch_1}$, $m_{\nt_2}$ limits for $m_{\nt_1} > 300$ GeV; 
For sleptons lighter than $\ch_1$ and $\nt_2$, 
the mass limits are $m_{\ch_1}, m_{\nt_2} > 1150$ GeV for $m_{\nt_1} \lesssim 700$ GeV and 
there is no $m_{\ch_1}$, $m_{\nt_2}$ limits for $m_{\nt_1} > 700$ GeV.

\item
The constraint on ($m_{A^0, H^+}, \tan\beta$) (at 95\% CL) from searches for the MSSM Higgs bosons 
$H^0$, $A^0$ and $H^+$ at LHC,~\cite{ICHEP2016_ATLAS,Charged_Higgs@ATLAS,SUSY@EPS-HEP2017,MSSM_Higgs@CMS}, 
where $H^0$ is the heavier CP-even Higgs boson.
\end{itemize}
 
\end{appendix}

%


\begin{thebibliography}{99}

\bibitem{Aad:2012tfa}
  G.~Aad {\it et al.} [ATLAS Collaboration],
  Phys. Lett. B  716 (2012) 1
  [arXiv:1207.7214 [hep-ex]].

\bibitem{Chatrchyan:2012xdj}
  \mbox{S.~Chatrchyan {\it et al.} [CMS Collaboration],
  Phys. Lett. B 716 (2012) 30} 
  [arXiv:1207.7235 [hep-ex]].

\bibitem{LHCcrosssecs}
  S.~Heinemeyer {\it et al.}  [LHC Higgs Cross Section Working Group Collaboration],
  arXiv:1307.1347 [hep-ph].

\bibitem{Djouadi:2005gi}
  A.~Djouadi,
  Phys.\ Rept.\  {\bf 457} (2008) 1
  [arXiv:hep-ph/0503172].

\bibitem{Bartl:2014bka}
  A.~Bartl, H.~Eberl, E.~Ginina, K.~Hidaka and W.~Majerotto,
  Phys. Rev. D 91 (2015) 015007
  [arXiv:1411.2840 [hep-ph]].

\bibitem{Eberl:h2bb}
  H.~Eberl, E.~Ginina, A.~Bartl, K.~Hidaka and W.~Majerotto
  JHEP 1606 (2016) 143 [arXiv:1604.02366 [hep-ph]].

\bibitem{QCD_corr}
  For the QCD corrections to $h^0 \to g g$, see 
  K. G. Chetyrkin, B. A. Kniehl, and M. Steinhauser, Nucl. Phys. B 510 (1998) 61 [arXiv:hep-ph/9708255]; 
  M. Kramer, E. Laenen, and M. Spira, Nucl. Phys. B511, 523 (1998), [arXiv:hep-ph/9611272];
  Y. Schroder and M. Steinhauser, JHEP 01 (2006) 051 , [arXiv:hep-ph/0512058];
  K. G. Chetyrkin, J. H. Kuhn, and C. Sturm, Nucl. Phys. B744 (2006) 121, [arXiv:hep-ph/0512060];
  P. A. Baikov and K. G. Chetyrkin, Phys. Rev. Lett. 97 (2006) 061803, [arXiv:hep-ph/0604194];
  T. Inami, T. Kubota, and Y. Okada, Z. Phys. C18 (1983) 69;
  A. Djouadi, M. Spira, and P. M. Zerwas, Phys. Lett. B264 (1991) 440;
  K. G. Chetyrkin, B. A. Kniehl, and M. Steinhauser, Phys. Rev. Lett. 79 (1997) 353 [arXiv:hep-ph/9705240];
  C. Anastasiou, C. Duhr, F. Dulat, F. Herzog, and B. Mistlberger, Phys. Rev. Lett. 114 (2015) 212001 
  [arXiv:1503.06056[hep-ph]];
  M. Schreck and M. Steinhauser, Phys. Lett. B655, 148 (2007) [arXiv:0708.0916 [hep-ph]].\\
  \\
  For the QCD corrections to $h^0 \to \gamma \gamma$, see 
  A. Djouadi, M. Spira, and P. M. Zerwas, Phys. Lett. B311 (1993) 255, [arXiv:hep-ph/9305335];
  H.-Q. Zheng and D.-D. Wu, Phys. Rev. D42 (1990) 3760;
  A. Djouadi, M. Spira, J. J. van der Bij, and P. M. Zerwas, Phys. Lett. B257 (1991) 187;
  S. Dawson and R. P. Kauffman, Phys. Rev. D47 (1993) 1264;
  K. Melnikov and O. I. Yakovlev, Phys. Lett. B312 (1993) 179 [arXiv:hep-ph/9302281];
  M. Inoue, R. Najima, T. Oka, and J. Saito, Mod. Phys. Lett. A9 (1994) 1189;
  J.Fleischer,O.V.Tarasov,andV.O.Tarasov, Phys. Lett. B584 (2004) 294 [arXiv:hep-ph/0401090];
  P. Maierh\"ofer and P. Marquard, Phys. Lett. B721 (2013) 131 [arXiv:1212.6233 [hep-ph]].

\bibitem{EW_corr}
  U. Aglietti, R. Bonciani, G. Degrassi, and A. Vicini, Phys. Lett. B595 (2004) 432, [arXiv:hep-ph/0404071];
  U. Aglietti, R. Bonciani, G. Degrassi, and A. Vicini, Phys. Lett. B600 (2004) 57, [arXiv:hep-ph/0407162];
  S. Actis, G. Passarino, C. Sturm, and S. Uccirati, Nucl. Phys. B811 (2009) 182, [arXiv:0809.3667[hep-ph]];
  G. Degrassi and F. Maltoni, Phys. Lett. B600 (2004) 255, [arXiv:hep-ph/0407249];
  S. Actis, G. Passarino, C. Sturm, and S. Uccirati, Phys. Lett. B670 (2008) 12, [arXiv:0809.1301[hep-ph]];
  G. Degrassi and F. Maltoni, Nucl. Phys. B724 (2005) 183, [arXiv:hep-ph/0504137].

\bibitem{Brignole:2015kva}
  A.~Brignole,
  Nucl. Phys. B 898 (2015) 644
  [arXiv:1504.03273 [hep-ph]].
  
\bibitem{Spira}
M. Spira, A. Djouadi, D. Graudenz and P. M. Zerwas, Nucl. Phys. B 453 (1995) 17 [arXiv:hep-ph/9504378].

\bibitem{Nojiri_Boselli}
M. Endo, T. Moroi, M. Nojiri, JHEP 04 (2015) 176 [arXiv:1502.03959 [hep-ph]]; 
S. Boselli, R. Hunter and A. Mitov, arXiv:1805.12027 [hep-ph].

\bibitem{Allanach:2008qq}
  B.~C.~Allanach {\it et al.},
  Comput.\ Phys.\ Commun.\  {\bf 180} (2009) 8
  [arXiv:0801.0045 [hep-ph]].
  
\bibitem{Gabbiani:1996hi}
  F.~Gabbiani, E.~Gabrielli, A.~Masiero and L.~Silvestrini,
  Nucl. Phys. B 477 (1996) 321
  [arXiv:hep-ph/9604387].

\bibitem{PDG2016}
  C. Patrignani et al. (Particle Data Group), Chin. Phys. C, 40, 100001 (2016).

\bibitem{Dedes}
A. Dedes et al., JHEP 2014 (2014) 137 [arXiv:1409.6546 [hep-ph]].

\bibitem{SPheno1}  
  W.~Porod, Comput. Phys. Commun. {\bf 153} (2003) 275 [arXiv:hep-ph/0301101].  

\bibitem{SPheno2}  
  W.~Porod and F.~Staub, Comput. Phys. Commun. {\bf 183} (2012) 2458 [arXiv:1104.1573
  [hep-ph]].
 
\bibitem{Pierce}  
D. M. Pierce et al., Nucl. Phys. B 491 (1997) 3.

\bibitem{Almeida:2013jfa}
  L.~G.~Almeida, S.~J.~Lee, S.~Pokorski and J.~D.~Wells,
  Phys. Rev. D 89 (2014) no.3,  033006
  [arXiv:1311.6721v3 [hep-ph]].

\bibitem{CERN_YR4}
LHC Higgs Cross Section Working Group, D. de Florian, C. Grojean, F. Maltoni, C. Mariotti,
A. Nikitenko, M. Pieri, P. Savard, M. Schumacher, R. Tanaka (Eds.), Handbook of LHC Higgs Cross
Sections: 4. Deciphering the nature of the Higgs sector, CERN Yellow Reports: Monographs,
Vol. 2/2017, CERN-2017-002-M (CERN, Geneva, 2017), https://doi.org/10.23731/CYRM-2017-002,
[arXiv:1610.07922].

\bibitem{ILC250}
K. Fujii et al., arXiv:1710.07621.

\bibitem{CLICref}
H. Abramowicz et al., arXiv:1307.5288[hep-ex].

\bibitem{ATLAS_kappa_plot}
ATLAS collaboration, ATLAS-CONF-2018-031.

\bibitem{CMS_kappa_plot}
CMS collaboration, CMS-HIG-16-040, CERN-EP-2018-060 [arXiv:1804.02716].

\bibitem{dga/g_error}
J. Tian, email correspondence.

\bibitem{Eberl_17}
H. Eberl, E. Ginina, K. Hidaka, Euro Physical Journal C77 (2017) 189 [arXiv:1702.00348 [hep-ph]].

\bibitem{tth@LHC}
W. Peng et al., Phys. Lett. B618 (2005) 209 
[arXiv:hep-ph/0505086 [hep-ph]]; 
S. Dittmaier et al., Phys. Rev. D90 (2014) 035010 
[arXiv:1406.5307 [hep-ph]].

\bibitem{PDG2019} 
M. Tanabashi et al. (Particle Data Group), Phys. Rev. D 98 (2018) 030001 and 2019 update.

\bibitem{epsK_DMK_SM}

J. Brod and M. Gorbahn, Phys. Rev. Lett. 108 (2012) 121801 [arXiv:1108.2036 [hep-ph]].

\bibitem{HFAG2016} 
  Y. Amhis et al. (Heavy Flavour Averaging Group), "Averages of b-hadron, 
   c-hadron, and tau-lepton properties as of summer 2016", arXiv:1612.07233[hep-ex].

\bibitem{DeltaMBs_SM}
  T. Jubb, M. Kirk, A. Lenz, and G. Tetlalmatzi-Xolocotzi, arXiv:1603.07770 [hep-ph];
  M. Artuso, G. Borissov, and A. Lenz, Rev. Mod. Phys. 88 (2016) 045002 
  [arXiv:1511.09466 [hep-ph]].

\bibitem{Misiak_2015}
  M. Misiak et al., Phys. Rev. Lett. 114 (2015) 221801 [arXiv:1503.01789[hep-ph]].

\bibitem{bsll_BABAR_2014}
  J.P. ~Lees {\it et al.}  [BABAR Collaboration],
  Phys. Rev. Lett. 112 (2014) 211802 [arXiv:1312.5364 [hep-ex]].

\bibitem{Huber_2008}
  T.~Huber, T.~Hurth and E.~Lunghi,
  Nucl. Phys. B 802 (2008) 40 [arXiv:0712.3009 [hep-ph]].

\bibitem{Bsmumu_LHCb_CMS}
  V. Khachatryan et al. [CMS and LHCb Collaborations],
  Nature 522 (2015) 68 [arXiv:1411.4413[hep-ex]].

\bibitem{Bsmumu_SM_Bobeth_2014}
  C. ~Bobeth {\it et al.},
  Phys. Rev. Lett. 112 (2014) 101801 [arXiv:1311.0903 [hep-ph]].

\bibitem{Trabelsi_EPS-HEP2015}
  K. ~Trabelsi, plenary talk at European Physical Society Conference on High Energy 
  Physics 2015 (EPS-HEP2015), Vienna, 22 - 29 July 2015.

\bibitem{Hamer_EPS-HEP2015} 
   P. ~Hamer, talk at European Physical Society Conference on High Energy 
   Physics 2015 (EPS-HEP2015), Vienna, 22 - 29 July 2015.

\bibitem{Btotaunu_LP2013}
 J. M. Roney, talk at 26th International Symposium on Lepton Photon
 Interactions at High Energies, San Francisco, USA, 24-29 June 2013.

\bibitem{Higgs_mass_ATLAS_CMS} 
 ATLAS and CMS collaborations, Phys. Rev. Lett. 114 (2015) 191803, 
  [arXiv:1503.07589[hep-ex]].

\bibitem{Higgs_mass_Heinemeyer}
 S. Borowka, T. Hahn, S. Heinemeyer, G. Heinrich and W. Hollik, 
  Eur. Phys. J. C75 (2015) 424 [arXiv:1505.03133 [hep-ph]].

\bibitem{kappa_b_ATLAS}
 H. M. Gray, talk at 54th Rencontres de Moriond on Electroweak Interactions 
 and Unified Theories, La Thuile, Italy, 16 - 23 Mar 2019 [ATLAS Collaboration, ATLAS-CONF-2019-005].

\bibitem{kappa_b_CMS}
 CMS Collaboration, Eur. Phys. J. C 79 (2019) 421 [arXiv:1809.10733 [hep-ex]].

\bibitem{SUSY@EPS-HEP2017}
  M. D'Onofrio, plenary talk at EPS Conference on High Energy Physics (EPS-HEP2017), 
  Venice, Italy, 5-12 July 2017.

\bibitem{SUSY@Moriond2017}
E. Kuwertz, talk at 52nd Rencontres de Moriond EW 2017, La Thuile, 18-25 March, 2017; 
  ATLAS Collaboration, ATLAS-CONF-2017-022 (submitted to 52nd Rencontres de Moriond on 
  Electroweak Interactions and Unified Theories, La Thuile, Italy, 18 - 25 Mar 2017); 
  A. Petridis, talk at 52nd Rencontres de Moriond on Electroweak Interactions and Unified 
  Theories, La Thuile, Italy, 18 - 25 Mar 2017.

\bibitem{slepton@LHC}
ATLAS Collaboration, JHEP 05 (2014) 071 [arXiv:1403.5294[hep-ex]];
  CMS Collaboration, EPJC 74 (2014) 3036 [arXiv:1405.7570[hep-ex]].

\bibitem{EWino@Moriond2017}
M. Marionneau, talk at 52nd Rencontres de Moriond on Electroweak Interactions and Unified Theories, 
  La Thuile, Italy, 18 - 25 Mar 2017.
  
\bibitem{Strandberg18}
S. Strandberg, plenary talk at XXXIX International Conference on High Energy
Physics, July 4-11, 2018, Seoul.

\bibitem{ICHEP2016_ATLAS}
 D. ~Charlton, plenary talk at 38th International Conference on High Energy Physics (ICHEP2016),
 Chicago, 3 - 10 August 2016.

\bibitem{Charged_Higgs@ATLAS} 
  C. Gwilliam, talk at 38th International Conference on High Energy Physics (ICHEP2016),
  Chicago, 3 - 10 August 2016;
  ATLAS Collaboration, ATLAS-CONF-2016-088.

\bibitem{MSSM_Higgs@CMS}
D. N. Taylor, talk at 54th Rencontres de Moriond on Electroweak Interactions 
 and Unified Theories, La Thuile, Italy, 16 - 23 Mar 2019; 
 CMS Collaboration, JHEP 1809 (2018) 007 [arXiv:1803.06553 [hep-ex]].

\bibitem{PDG2018}
  M. Tanabashi et al. (Particle Data Group), Phys. Rev. D 98, 030001 (2018).


\end{thebibliography}
\end{document}